\documentclass[aps,prxquantum,floatfix,twocolumn,superscriptaddress]{revtex4-2}
\usepackage[utf8]{inputenc}
\usepackage{amsmath}
\usepackage{amssymb}
\usepackage{listings}
\usepackage{graphicx,subcaption}
\usepackage{dcolumn}
\usepackage[export]{adjustbox}
\usepackage{color, soul, cancel} 
\usepackage[export]{adjustbox}
\usepackage{braket}
\usepackage{natbib}
\usepackage{orcidlink}
\usepackage{appendix}
\usepackage{physics}
\usepackage{mathrsfs}
\usepackage[T1]{fontenc}
\usepackage{lmodern}
\usepackage{latexsym}
\usepackage{graphicx}
\usepackage{rotating}
\usepackage[normalem]{ulem}
\usepackage{hyperref}
\usepackage{amsmath,amsfonts}
\usepackage{bm}
\usepackage{xcolor}
\usepackage{float}
\usepackage{textgreek}
\usepackage{qcircuit}
\usepackage[greek,english]{babel}

\usepackage[justification=justified,singlelinecheck=false]{caption}
\usepackage{ragged2e}
\hypersetup{colorlinks=true,urlcolor=blue,linkcolor=blue}

\begin{document}
\title{Quantum Assisted Ghost Gutzwiller Ansatz}
\author{P.V. Sriluckshmy\orcidlink{0009-0009-5545-5954}}
\affiliation{IQM, Nymphenburgerstr. 86, 80636 Munich, Germany}
\author{François Jamet}
\affiliation{IQM, Nymphenburgerstr. 86, 80636 Munich, Germany}
\author{Fedor \v{S}imkovic IV\orcidlink{0000-0003-0637-5244}}
\affiliation{IQM, Nymphenburgerstr. 86, 80636 Munich, Germany}
\begin{abstract}
 The ghost Gutzwiller ansatz (gGut) embedding technique was shown to achieve comparable accuracy to the gold standard dynamical mean-field theory method in simulating real material properties, yet at a much lower computational cost. Despite that, gGut is limited by the algorithmic bottleneck of computing the density matrix of the underlying effective embedding model, a quantity which must be converged within a self-consistent embedding loop. We develop a hybrid quantum-classical gGut technique which computes the ground state properties of embedding Hamiltonians with the help of a quantum computer, using the sample-based quantum-selected configuration interaction (QSCI) algorithm. We study the applicability of SCI-based methods to the evaluation of the density of states for single-band Anderson impurity models within gGut and find that such ground states of interest become sufficiently sparse in the CI basis as the number of ghost orbitals is increased. Further, we investigate the performance of QSCI using local unitary cluster Jastrow (LUCJ) variational quantum states in combination with a circuit cutting technique, prepared on IQM's quantum hardware for system sizes of up to 11 ghost orbitals, equivalent to 24 qubits. We report converged gGut calculations which correctly capture the metal-to-insulator phase transition in the Fermi-Hubbard model on the Bethe lattice by using quantum samples to build an SCI basis with as little as $1\%$ of the total CI basis states.
\end{abstract}
\date{\today}
\maketitle
\section{Introduction}\label{sec:intro}

The accurate description of the physical properties of materials has proven to be a formidable challenge due to the exponentially growing computational space and the infamous sign problem associated with the fermionic nature of electronic interactions \cite{troyer2005computational}. A number of embedding methods have been designed with the idea of approximating such problems with simpler effective models, which couple a small fragment of the original Hamiltonian to an effective bath, the properties of which have to be determined self-consistently. 

Such approaches are controlled in the sense that the exact result can be recovered by gradually increasing the fragment and bath sizes~\cite{georges1996dynamical,maier2005quantum}. In practice, however, the range of applicability of currently available numerical \emph{ impurity solvers} designed to solve such effective models is limited to only small systems, which is insufficient to extract accurate properties for many materials of interest. With the advent of quantum computing, there is hope that classical impurity solvers may eventually be surpassed by their quantum counterparts. Despite a number of studies that focus on hybrid quantum-classical embedding methods \cite{yamazaki2018towards,jamet2021krylov,ma2021quantum,yao2021gutzwiller,besserve2022unraveling,greene2022modelling,iijima2023towards,dhawan2024quantum,bertrand2024turning,shajan2024towards,jamet2025anderson,ehrlich2025variational,sheng2022green,rusakov2018self,lan2017generalized,tilly_2021}, it is an open question which combination of classical embedding theory and quantum algorithm holds the greatest promise \cite{sun2016quantum,vorwerk2022quantum}.

The dynamical mean-field theory (DMFT) method \cite{georges1996dynamical} and extensions thereof \cite{hettler2000dynamical,kotliar2001cellular,maier2005quantum,rohringer2018diagrammatic,klett2020real} have been immensely successful in describing the properties of solid-state systems defined on periodic lattices, in particular in combination with ab-initio approaches such as density function theory (DFT) \cite{pavarini2011lda,held2001realistic} and the GW method \cite{sun2002extended,boehnke2016strong,zhu2021ab}. Despite being widely adopted as the go-to algorithm in such settings, there are a number of complexities within DMFT that currently prevent sufficiently scalable implementations of the algorithm for many materials of scientific interest. 

For quantum Monte Carlo (QMC) impurity solvers within DMFT \cite{gull2011continuous}, computations must be performed at finite temperature and in imaginary time, which requires the application of an ill-defined analytic continuation procedure to rotate the results back to the real axis for experimental comparisons~\cite{vidberg1977solving,jiani2021nevanlinna}. For other solvers, such as the exact diagonalization method (ED) \cite{rozenberg1994metal,caffarel1994exact,georges1996dynamical,zgid2012truncated,lin2013efficient}, the DMFT bath must be discretized, which limits the applicability of the algorithm as the computational complexity grows exponentially with the system size. Tensor networks (TN) represent yet another alternative, but suffer from the rapidly growing bond dimension when performing the time-evolution of a quantum state, a necessary step in the computation of the single-particle Green's function, essential to the DMFT formalism~\cite{ganahl2015efficient, bauernfeind2017fork, cao2021tree, cao2024finite}.

Since none of the aforementioned solvers offers a truly scalable solution for DMFT, alternative embedding methods with lower algorithmic complexity have been introduced, notably the density matrix embedding theory (DMET)\cite{knizia2012density,knizia_2013,wouters_2016} and the Gutzwiller ansatz (GA)\cite{lanata2012efficient} also known under the name of rotationally-invariant slave bosons (RISB)\cite{lanata2017critical,lanata2019connection,ayral2017dynamical}. These methods do not require the direct computation of the Green's function and instead rely on the one-particle density matrix (1-RDM) as their central quantity, which has to be converged within a self-consistent loop. Moreover, the baths in both methods are discrete by construction, such that one bath orbital is defined for every orbital of the impurity. However, these algorithmic simplifications are enabled through additional approximations, leading to less accurate results compared to DMFT. In some cases, this can even result in qualitatively inaccurate descriptions of underlying physics, i.e. the failure to capture the Mott (metal-to-insulator) transition  in transition-metal oxides\cite{lee2023accuracy_single}. 

Several algorithms have been developed with the aim of interpolating between DMFT and DMET/GA while striking the right balance between performance and the precision of results \cite{lanata2023derivation,sriluckshmy_2021}. The ghost Gutzwiller Ansatz (gGut)\cite{lanata2022operatorial,lee2023accuracy_single, lee2023accuracy_multi, mejuto2023efficient,lee2024charge,frank2024active,tagliente2025revealing} is one such example, defined as the extension of GA wherein the number of bath orbitals (ghosts) can be systematically increased. The gGut method preserves the advantages of computational simplicity provided by GA, while being variational,  and converges to DMFT in the infinite-bath limit \cite{ayral2017dynamical}. gGut performs well in capturing the complex behavior of single- and multi-orbital lattice models \cite{lee2023accuracy_single,lee2023accuracy_multi}, as well as molecules \cite{mejuto2024quantum}, even with a relatively small number of ghosts. Despite the fact that the method has been applied to three-band models with up to 36 fermionic modes using the TN-related density matrix renormalization group (DMRG) impurity solver \cite{white1992density, zhai2023block2, lanata2023derivation}, the question remains of how far it can be pushed using classical methods. At the same time, gGut is an ideal playground for early quantum computing algorithms due to the relative sparseness of the embedding Hamiltonian, reasonable qubit requirements, and the absence of the need to implement deep time-evolution circuits as required for the computation of the Green's function. 

In this paper, we assess the applicability of classical and quantum algorithms based on the selected configuration interaction (SCI) methodology \cite{huron1973iterative, ohtsuka2017selected,levine2020casscf} as impurity solvers within gGut. In particular, we focus on the recently popularized QSCI \cite{kanno2023quantum, nakagawa2024adapt, mikkelsen2024quantum} algorithm, also referred to as quantum subspace diagonalization (SQD)~\cite{ieva2025quantum,yu2025sample}, which computes ground state properties by diagonalizing the Hamiltonian of interest in the basis of computational basis samples obtained from a quantum trial state. For our hybrid quantum-classical gGut workflow, we use trial states based on the parameterized local unitary cluster Jastrow (LUCJ) ansatz \cite{motta2023bridging}, which we prepare on IQM's quantum hardware \cite{abdurakhimov2024technology} using a circuit cutting technique \cite{peng2020simulating,lowe2023fast}. This allows us to compute the density of states for the Fermi-Hubbard model on the Bethe lattice for systems of up to 24 qubits and reliably capture the metal-to-insulator transition in the model as a function of interaction strength.  

The paper is structured as follows: in Section~\ref{sec:method} we provide an introduction to the gGut method, Section~\ref{sec:hybrid_workflow} is dedicated to the description of our quantum computing workflow, we present our classical and quantum results in Section~\ref{sec:classical} and Section~\ref{sec:quantum}, respectively, and provide a summary and outlook in Section~\ref{sec:conclusions}.
\section{Ghost Gutzwiller Ansatz}\label{sec:method}
\begin{figure*}
\includegraphics[width=0.95\textwidth]{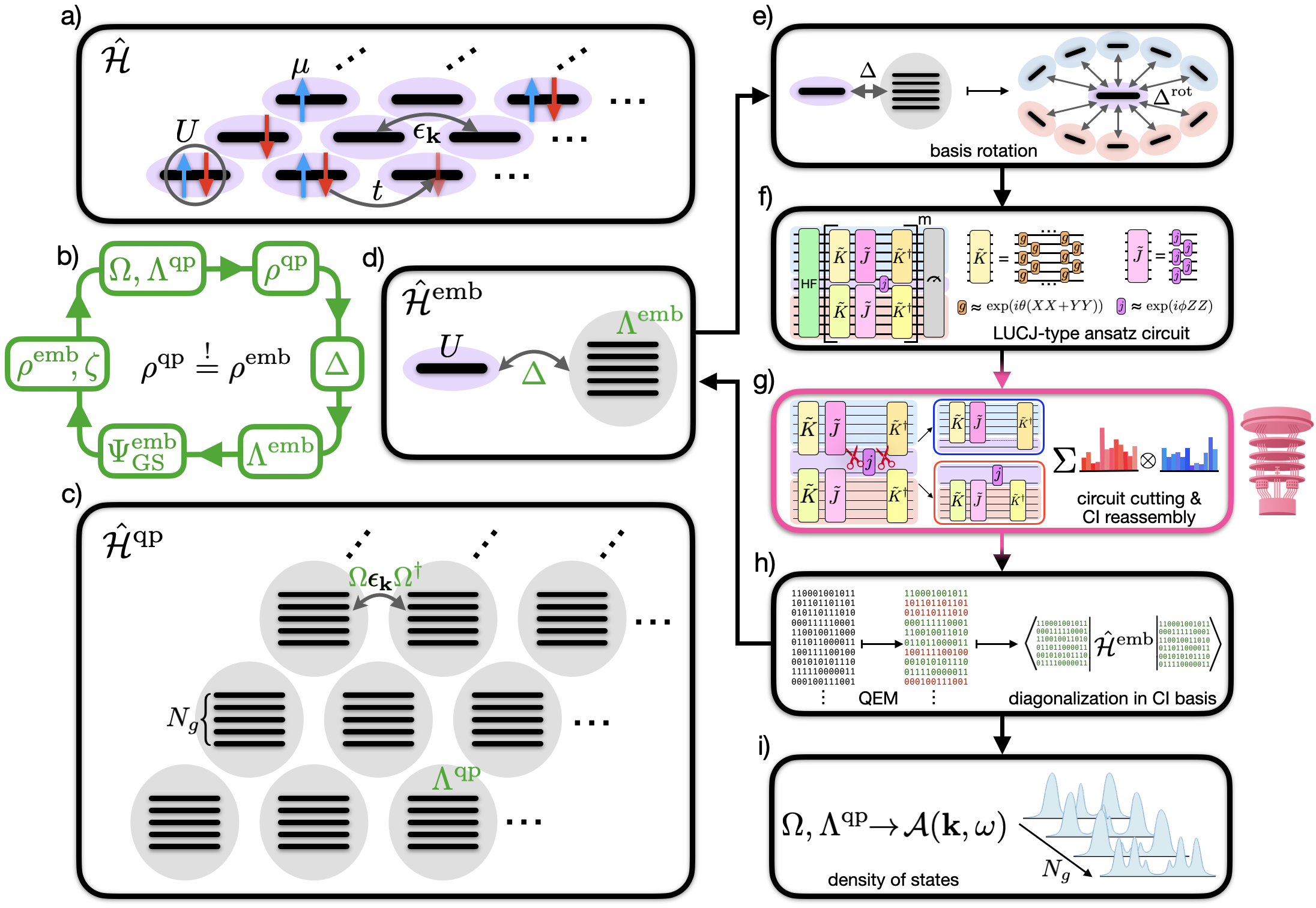}
\caption{\justifying \textbf{gGut: classical and quantum workflow.} Instead of attempting to directly solve the system of interest defined on a periodic lattice (a), a self-consistent loop (b) involving a quasi-particle Hamiltonian (c) and an embedding Hamiltonian (d) is executed until the density matrices of both models have converged to the same value. In our hybrid quantum-classical workflow, the ground-state density matrix of the embedding Hamiltonian is computed through the following steps: First, a single-particle basis transformation of the Hamiltonian is performed (e). Then, an LUCJ trial state (f) is prepared, on IQM's quantum hardware using a circuit cutting technique (g), where CI samples are reassembled as tensor products of measurements from multiple cut circuits. Finally, an error mitigation procedure is applied, filtering out samples in wrong particle sectors and the embedding Hamiltonian is diagonalized in the basis of the remaining CIs. Upon convergence of the gGut algorithm, dynamical quantities such as the density of states can be computed from the knowledge of the model parameters from the quasi-particle Hamiltonian (i).}
\label{fig:main}
\end{figure*}

In this section, we introduce the Ghost Gutzwiller Ansatz (gGut) algorithm \cite{lanata2022operatorial} laying the foundation for our hybrid quantum-classical workflow. The constituent parts of this workflow are represented graphically in Fig.~\ref{fig:main}. Our model of interest throughout this work will be the half-filled single-band Fermi-Hubbard model defined on the Bethe lattice \cite{eckstein2005hopping} (see Fig.~\ref{fig:main}a). However, the methodology is not restricted to this example and can be equally applied to other more complex models from solid-state physics\cite{lanata2022operatorial} and quantum chemistry\cite{mejuto2024quantum}. The Hamiltonian of the Fermi-Hubbard model is given by:
\begin{align}
    \hat{\mathcal{H}} = \sum_{\mathbf{k}\sigma} \epsilon_{\mathbf{k}\sigma} \hat{c}^{\dagger}_{\mathbf{k}\sigma} \hat{c}^{\phantom{\dagger}}_{\mathbf{k}\sigma} + U \sum_{\mathbf{r}} \hat{n}_{\mathbf{r}\uparrow}^{\phantom{\dagger}} \hat{n}_{\mathbf{r} \downarrow}^{\phantom{\dagger}} + 
    \mu \sum_{\mathbf{r}\sigma} \hat{n}_{\mathbf{r}\sigma}^{\phantom{\dagger}}
\end{align}
where $\hat{c}^{\dagger}_{\mathbf{k}\sigma}$($\hat{c}_{\mathbf{k}\sigma}^{\phantom{\dagger}}$) creates(annihilates) a fermion of momentum $\mathbf{k}$ and spin $\sigma \in \{\uparrow, \downarrow\}$, $\hat{n}_{\mathbf{r}\sigma}^{\phantom{\dagger}} = \hat{c}^{\dagger}_{\mathbf{r}\sigma} \hat{c}_{\mathbf{r}\sigma}$ counts the number of fermions of spin $\sigma$ on lattice site $\mathbf{r}$, $\epsilon_{\mathbf{k}}$ is the dispersion relation, $U$ is the repulsive on-site interaction strength, and $\mu$ is the chemical potential used to adjust the number of particles in the system. The density of states for the Fermi-Hubbard model on the infinite Bethe lattice corresponds to the semi-circular function $\mathcal{A}
_0(\omega) = \frac{1}{\pi t} \sqrt{1-(\frac{\omega}{2t})^2}$, where $\omega$ is the frequency and $t$ is the nearest-neighbor hopping amplitude. Following standard practice, the energy unit is set to be the half-bandwidth $D=2t$ throughout this work. 

Instead of directly tackling the Hamiltonian $\hat{\mathcal{H}}$, the gGut algorithm iteratively evaluates the quasi-particle and embedding Hamiltonians $\hat{\mathcal{H}}^{\text{qp}}$ and $\hat{\mathcal{H}}^{\text{emb}}$ in a self-consistent loop. Here, $\hat{\mathcal{H}}^{\text{qp}}$ serves as a parameterized low-level description of the lattice system,  while $\hat{\mathcal{H}}^{\text{emb}}$ serves as a parameterized high-level description of a local system, and the parameters of the two Hamiltonians are tied to each other through a set of self-consistent conditions.
While $\hat{\mathcal{H}}^{\text{qp}}$ is quadratic and can be readily diagonalized, it is non-trivial to obtain the ground state $\Psi^{\text{emb}}_{\text{GS}}$ of $\hat{\mathcal{H}}^{\text{emb}}$, and its evaluation using a classical or quantum impurity solver constitutes the computational bottleneck of gGut. Let us now proceed to introduce the two effective Hamiltonians used in gGut in more detail.

The quasi-particle Hamiltonian (see Fig.~\ref{fig:main}c) is given by:
 \begin{align}
     \hat{\mathcal{H}}^{\text{qp}} = \sum_{\mathbf{k}ab\sigma} \left( \Omega^{\phantom{\dagger}}_{a\sigma} \epsilon_{\mathbf{k}\sigma} \Omega^{\dagger}_{b\sigma} + \Lambda^{\text{qp}}_{ab\sigma} \right) \hat{f}^{\dagger}_{\mathbf{k}a\sigma}\hat{f}^{\phantom{\dagger}}_{\mathbf{k}b\sigma}, \label{eq:qphamiltonian}
 \end{align}
where $\hat{f}^{\dagger}$ and $\hat{f}$ are the creation and annihilation operators for the fermionic ghost degrees of freedom with spin $\sigma$ and are indexed by the roman letters $a,b, \ldots \in \{1,2,\dots,N_g\}$ where $N_g$ is the total number of ghosts per spin orbital of the impurity. We note that, following standard practice, we only consider odd values of $N_g$ in this work \cite{frank2021quantum}. Further, $\Omega$ is the quasi-particle renormalization matrix and $\Lambda^{\text{qp}}$ is the renormalized 
potential.

The embedding Hamiltonian (see Fig.~\ref{fig:main}d) is defined as:
\begin{align}
    &\hat{\mathcal{H}}^{\text{emb}} \; = U  \hat{n}_{0\uparrow}^{d} \hat{n}_{0\downarrow}^{d} + \mu \sum_{\sigma} \hat{n}_{0\sigma}^{d} \\& 
    +\sum_{a\sigma}  \Delta_{a \sigma} \left(\hat{d}^{\dagger}_{0\sigma} \hat{d}^{\phantom{\dagger}}_{a\sigma}
    + \hat{d}^{\dagger}_{a\sigma} \nonumber \hat{d}^{\phantom{\dagger}}_{0\sigma} \right)
    - \sum_{ab\sigma} \Lambda^{\text{emb}}_{ba}  \hat{d}^{\dagger}_{a\sigma}\hat{d}^{\phantom{\dagger}}_{b\sigma}, \nonumber 
\end{align}
where $\hat{d}^{\dagger}_{0\sigma}$ and $\hat{d}_{0\sigma}$ are the creation and annihilation operators of the impurity orbitals, while $\hat{d}^{\dagger}_{a\sigma}$ and $\hat{d}_{a\sigma}$ are the corresponding ladder operators of the ghost orbitals $a,b, \ldots \in \{1,2,\dots,N_g\}$, and the corresponding density operators are defined as $\hat{n}_{i\sigma}^{d} = \hat{d}^{\dagger}_{i\sigma} \hat{d}_{i\sigma}^{\phantom{\dagger}}$. Further, $\Delta$ is the hybridization potential, which couples impurity orbitals to ghost orbitals, and $\Lambda^{\text{emb}}$ is the ghost potential.

Let us now focus on the self-consistent gGut loop, coupling the two effective Hamiltonians, as presented graphically in (see Fig.~\ref{fig:main}b). Through a set of equations, the model parameters of $\{\Omega, \lambda^{\text{qp}},\rho^{\text{qp}},\Delta, \Lambda^{\text{emb}},\Psi^{\text{emb}}_{\text{GS}}, \rho^{\text{emb}},\zeta \}$ (indicated by green color in Fig.~\ref{fig:main}) of $\hat{\mathcal{H}}^{\text{qp}}$ and $\hat{\mathcal{H}}^{\text{emb}}$ are determined in sequence. Here, $\rho^{\text{qp}}$ and $\rho^{\text{emb}}$ are the one-particle reduced density matrices (1-RDM) of  $\hat{\mathcal{H}}^{\text{qp}}$ and $\hat{\mathcal{H}}^{\text{emb}}$, respectively, defined on the ghost degrees of freedom, whilst $\zeta$ is the 1-RDM coupling the ghosts to the impurity in $\hat{\mathcal{H}}^{\text{emb}}$. The steps of the self-consistent loop are as follows (note that we suppress spin indices in all equations to improve readability):
 
\begin{enumerate}
    \item The algorithm begins with an initial guess for $\Omega$ and $\Lambda^{\text{qp}}$. From the knowledge of $\Omega$ and $\Lambda^{\text{qp}}$, the 1-RDM for the quasi-particle Hamiltonain $\rho^{\text{qp}}$ is computed using:
    \begin{align}
    \rho_{ab}^{\text{qp}} =& \frac{1}{\mathcal{N}} \left[\sum_{\mathbf{k}} \operatorname{n}_{F} \left( \Omega\epsilon_{\mathbf{k}} \Omega^{\dagger}_{} + \Lambda^{\text{qp}}\right) \right]_{ba} \label{eq:rhoqp}
    \end{align}
    where $n_{F}(x) = \frac{1}{e^{x/T} + 1}$ is the Fermi-Dirac distribution function at temperature $T$ and $\mathcal{N}$ is a normalization constant corresponding to the total number of lattice sites. Note that the expression in the square bracket is obtained by diagonalizing the quasi-particle Hamiltonian.

    \item Next, we compute $\Delta$ and $\Lambda^{\text{emb}}$ from:
    \begin{align}
        \Delta_{a} &= \\  &\frac{1}{\mathcal{N}} \sum_{\mathbf{k}c} [\rho^{\text{qp}}(1-\rho^{\text{qp}})]^{-1/2}_{ac}  \left[ \epsilon_{\mathbf{k}} \Omega^{\dagger} \operatorname{n}_{F}(\Omega \epsilon_{\mathbf{k}} \Omega^{\dagger} + \Lambda^{\text{qp}})\right]_{c} \nonumber \label{eq:Deltaqp}
    \end{align}
    \begin{align}     \Lambda^{\text{emb}}_{ab} &= -\sum_{cd} \left( \Omega_{c} \frac{\partial  \left[\rho^{\text{qp}}(1-\rho^{\text{qp}})\right]^{1/2}_{cd}  }{\partial\rho^{\text{qp}}_{ab}}\Delta_{d} + \text{c.c.} \right) -\Lambda^{\text{qp}}_{ab}. 
    \end{align}
    \item The updated values for $\Delta$ and $\Lambda^{\text{emb}}$  define a new embedding Hamiltonian, which must be solved numerically to obtain its ground state wavefunction $\Psi_{\text{GS}}^{\text{emb}}$, the corresponding ground state energy $E^{\text{emb}}$, and the 1-RDMs, $\rho^{\text{emb}}$ and $\zeta$ from:
    \begin{align}
        & \hat{\mathcal{H}}^{\text{emb}} \ket{\Psi_{\text{GS}}^{\text{emb}}} = E^{\text{emb}} \ket{\Psi_{\text{GS}}^{\text{emb}}} \\
    & \bra{\Psi_{\text{GS}}^{\text{emb}}} \hat{d}^{\dagger}_{0} \hat{d}^{\phantom{\dagger}}_{a} \ket{\Psi_{\text{GS}}^{\text{emb}}} = \zeta_{a} \\
    & \bra{\Psi_{\text{GS}}^{\text{emb}}} \hat{d}^{\dagger}_{a} \hat{d}^{\phantom{\dagger}}_{b}  \ket{\Psi_{\text{GS}}^{\text{emb}}} = \rho^{\text{emb}}_{ab}, 
    \end{align}
    \item Finally, new $\Omega$ and $\Lambda^{qp}$ values are computed from the parameters of the embedding model using
    \begin{align}
     \Omega_{a} &= \sum_{b} \zeta_{b} \left[\rho^{\text{emb}} (1-\rho^{\text{emb}})\right]^{-1/2}_{ba} \\
     \Lambda^{\text{qp}}_{ab}  &= -\sum_{cd} \left(\Omega_{c} \frac{\partial  \left[\rho^{\text{emb}}(1-\rho^{\text{emb}})\right]^{1/2}_{cd}  }{\partial\rho_{ab}} \Delta_{d} + \text{c.c.} \right) -\Lambda^{\text{emb}}_{ab},
    \end{align}
and the loop is repeated from the first step until a convergence of $|\Omega|$ , $|\Lambda^{qp}|$ and  $|\operatorname{Tr}(\rho^{\text{qp}}) - \operatorname{Tr}(\rho^{\text{emb}})|$ is observed within some stated tolerance limit. 
\end{enumerate}

Once the self-consistent loop has converged, the Green's function can be expressed as a function of $\Omega$ and $\Lambda^{\text{qp}}$, given by:
\begin{align}
    G (\mathbf{k},\omega) = \sum_{ab} \Omega_{a}^\dagger \left[ \omega + i0^+ - \Omega\epsilon_{\mathbf{k}} \Omega^\dagger -\Lambda^{\text{qp}} \right]_{ab}^{-1} \Omega_{b}. \label{eq:greensfunc}
\end{align}
The local density of states (DOS), a central quantity in the determination of material properties (see Fig.~\ref{fig:main}i), can then be computed from the imaginary part of the Green's function as:
\begin{align}
    \mathcal{A}(\omega) = -\frac{1}{\pi} \operatorname{Im} G(\omega).
\end{align}
Other relevant quantities for the determination of the properties of the system of interest, such as self-energy $\Sigma$ and quasi-particle weight $Z$ can be equally derived from knowledge of the parameters of the quasi-particle Hamiltonian $\hat{\mathcal{H}}^{\text{qp}}$~\cite{lee2023accuracy_single,lee2023accuracy_multi}.

\section{Hybrid Impurity Solver Workflow}\label{sec:hybrid_workflow}
\subsection{(Quantum) Selected Configuration Interaction}
\label{sec:qsci}

The computational (CI) basis encompasses all wavefunctions of the form:
\begin{align}\label{eq:basis}
    \ket{\Psi} = \sum_{x \in \mathcal{S}} \alpha_x \ket{x},
\end{align}
where $\mathcal{S}\equiv \{0,1\}^{N}$ is the set of all possible basis states, $N$ is the total number of spin-orbitals, the integers $\{0,1\}$ indicate whether orbitals are (un-)occupied, and the computational states $\ket{x}$ are Slater determinants, also called CI states. This CI basis includes the full configuration interaction (FCI)\cite{gao2024distributed} wavefunction corresponding to the ground state of a Hamiltonian of interest, $\ket{\Psi_{\text{FCI}}}=\ket{\Psi_{\text{GS}}}$. In order to determine the coefficients $\alpha_{x}$ of individual basis states in the FCI wavefunction, the Hamiltonian in a matrix form defined by:
\begin{align}
    \mathcal{H}_{xy} = \bra{x} \hat{\mathcal{H}} \ket{y} \quad x,y \in \mathcal{S},
\end{align}
is diagonalized in the CI basis of Eq.~\ref{eq:basis}:
\begin{align}
    \hat{\mathcal{H}} \ket{\Psi_{\text{FCI}}} = E_{\text{GS}} \ket{\Psi_{\text{FCI}}},
\end{align}
which can be performed in $O(\mathcal{S}_{\text{sym}}^3)$ computational steps. Here, $S_{\text{sym}} \subseteq \mathcal{S}$ is the set of basis states in the correct symmetry (i.e. particle number) sector, which grows exponentially with the system size $N$, thus limiting the applicability of this brute-force diagonalization approach to relatively small systems.

Various selected configuration interaction (SCI) methods have been developed with the purpose of investigating systems beyond the reach of FCI by exploiting the fact that the FCI wavefunction may be sparse in the computational basis, i.e. dominated by a relatively small fraction of all basis states. SCI algorithms are heuristic in nature and typically rely on iterative energetic~\cite{holmes2016heat} or multi-reference perturbative criteria~\cite{evangelisti1983convergence,loos2020the} for the selection of the subset of relevant CIs $\mathcal{S_{\text{SCI}}} \subset \mathcal{S_{\text{sym}}}$ where ideally $|\mathcal{S_{\text{SCI}}}|\ll |\mathcal{S_{\text{sym}}}|$. The performance of SCI methods relies on their ability to identify a compact subspace that effectively captures the essential electron correlation effects in the system, so that $\ket{\Psi_{\text{SCI}}} \simeq \ket{\Psi_{\text{FCI}}}$.

In the quantum selected configuration interaction (QSCI) approach, a quantum device is employed to prepare an initial quantum state, which is then measured in the computational basis to obtain the set of CI states $\mathcal{S}_{\text{SCI}}$. This procedure may yield an improvement over purely classical SCI methods, provided that the quantum state cannot be efficiently prepared using classical resources and that it performs better at capturing relevant CIs of the ground state wavefunction. We note that to achieve good QSCI performance, the quantum trial state itself does not necessarily need to have a high overlap with the ground state wavefunction, but rather should ideally be a uniform linear superposition of all relevant CI states required to compute an observable of interest up to a certain precision~\cite{reinholdt2025exposing}. However, in general, it is not known how to efficiently identify and prepare such states.

In this work, we use the SCI and QSCI methods to solve the embedding Hamiltonian $\hat{\mathcal{H}}^{\text{emb}}$ within the self-consistent loop of gGut. The choice of the size of the $\mathcal{S}_{\text{SCI}}$ will ultimately depend on a number of factors, such as the required precision of our observables of interest and on model parameters of the Hamiltonian $\hat{\mathcal{H}}^{\text{emb}}$.

An aspect with significant influence on the rate of convergence of observables is the choice of basis for the representation of the Hamiltonian \cite{motta2023bridging, yu2025sample} (see Fig.\ref{fig:main}e). In this work, the original single-particle basis is first rotated to a tri-diagonal basis, forming a chain configuration. A subsequent rotation to a diagonal/semi-diagonal basis is then performed, wherein one can choose the number of rotated orbitals. Specifically, the basis where all ghost sites are rotated is referred to as the star configuration. If all the  ghost orbitals and the impurity orbital are rotated together, one obtains the canonical basis. We will study the relative performance of different bases in the context of QSCI in the following section. 

The truncation of $\mathcal{S}_{\text{SCI}}$ within QSCI can be performed either by setting a cutoff for the individual CI weights in the ansatz state, fixing the total number of CIs, or stochastically resampling the CIs. For the DOS results obtained in this paper, we use the second option.

\subsection{LUCJ Quantum Trial State}
\label{sec:lucj}

Several types of quantum trial states have been explored in the context of QSCI in previous works, including adaptively constructed parameterized circuits and real-time-evolved Hartree-Fock states~\cite{kanno2023quantum, nakagawa2024adapt, mikkelsen2024quantum}. Other works have used a sample-based Krylov subspace technique \cite{ieva2025quantum} with local unitary cluster Jastrow (LUCJ) \cite{motta2024quantum} as the ansatz state. The LUCJ ansatz has been shown to provide good expressivity and perform especially well as a trial state for QSCI beyond the Hartree-Fock solution. In the LUCJ ansatz, an initial state is evolved using nearest-neighbor density-density terms surrounded by orbital rotations:
\begin{align}
    \ket{\Psi_{\text{LUCJ}}} \equiv \prod _{m}^{} e^{i K_{m}} e^{i J_{m}} e^{-i K_{m}}  \ket{\Psi_0}
\end{align}
with
\begin{align}
K_{m} & = \sum_{\langle i,j\rangle,\sigma} \kappa_{ij}^{m\sigma}  f_{i\sigma}^\dagger f_{j\sigma}^{\phantom{\dagger}} - h.c. \\
J_m &= \gamma^{m\uparrow\downarrow} \hat{n}_{0\uparrow}^f\hat{n}_{0\downarrow}^f +  \sum_{\langle i,j\rangle,\sigma}  \gamma^{m\sigma}_{ij} \hat{n}_{i\sigma}^f \hat{n}_{j\sigma}^f. 
\end{align}
where the sums run over nearest neighbors $\langle i,j\rangle$ of the same spin, and the values of tunable parameters $\kappa$ and $\gamma$ are typically initialized using the $T_2$ amplitudes of a coupled-cluster singles-doubles (CCSD) calculation~\cite{bartlett2007coupled}. The full sequence (layer) is repeated $m$ times within the ansatz circuit, each with different parameters for $\kappa$ and $\gamma$. Under the Jordan-Wigner transformation, used as the fermion-to-qubit mapping of choice throughout this work, these terms take the form:
\begin{align}
    \ket{\Psi_{\text{LUCJ}}^{\text{JW}}} = \prod_{m} \tilde{K}_{m}^{\phantom{\dagger}} \tilde{J}_{m}^{\phantom{\dagger}} 
    \tilde{J}_{m\uparrow \downarrow}^{\phantom{\dagger}} 
    \tilde{K}_{m}^{\dagger} \ket{\Psi_0}
\end{align}
where
\begin{align}
      \tilde{K}_{m} &\equiv \prod_{\langle i,j\rangle,\sigma}e^{i \theta_{ij}^{m\sigma} (X_i^{\sigma}X_j^{\sigma}+Y_i^{\sigma}Y_j^{\sigma})} \prod_{i,\sigma}e^{i \lambda_{i}^{m\sigma} P_i^{\sigma}} \\
      \tilde{J}_{m} &\equiv \prod_{\langle i,j\rangle,\sigma} e^{i \phi_{ij}^{m\sigma} Z_i^{\sigma} Z_j^{\sigma}}\\
      \tilde{J}_{m}^{\uparrow \downarrow} &\equiv e^{i \phi_{ij}^{m\uparrow\downarrow} Z_0^{\uparrow} Z_0^{\downarrow}},
\end{align}
where $\{\theta,\lambda,\phi\}$ are tunable parameters determined from $\kappa$ and $\gamma$. In this ansatz, we restrict spin up-down interactions to the impurity site, in line with the structure of $\hat{\mathcal{H}}^{\text{emb}}$. A pictorial representation of the ansatz circuit, such that the qubits corresponding to the impurity orbitals are placed at the center of the circuit, is shown in Fig.~\ref{fig:main}f. The number of parameters to be optimized grows as $m N(N+2)/4$ where $N = 2(N_o + N_g)$ is the total number of spin-orbitals of the system that contains $N_o$ impurity orbitals and $N_g$ ghost orbitals. When applying the $\exp(i \theta_{ij}^m (X_i X_j+Y_i Y_j))$ orbital rotation gates to single particle configuration states, some gates will leave the particle state unchanged irrespective of the parameters and can be removed, thereby reducing the total number of parameters in the circuit~\cite{motta2023bridging}. The circuit depth can be also further reduced by combining orbital rotations from subsequent layers~\cite{motta2023bridging}.  
\subsection{Circuit Cutting}
\label{sec:circuitcutting}

Circuit cutting techniques have been introduced with the aim of reducing quantum hardware requirements in terms of the number of qubits as well as the total number of entangling gates to be executed within a quantum circuit~\cite{peng2020simulating}. Typically, a quantum channel $\mathcal{V}$ is decomposed using $\mathcal{V} = \sum_i a_i \mathcal{F}_i$ where $a_i$ are real numbers. For gate cutting ~\cite{mitarai2021constructing} the decomposed channel is non-local and unitary and $\mathcal{F}_i$ are local unitary/measurement channels. For wire cutting \cite{brenner2023optimal} the decomposed channel is the identity and $\mathcal{F}_i$ are prepare/measurement channels. In this work, the Pauli wire cutting method is used, in which the density matrix of a single qubit is expressed as a linear combination in the orthogonal basis of single-qubit operators \cite{peng2020simulating}:
\begin{align}
    \rho & = \frac{1}{2} \sum_i c_i \operatorname{Tr} (\rho\, O_i) \rho_i \label{eq:rho}
\end{align}
where the sum runs over the operator matrices $O_i$ including the identity operator $I$, and the Pauli operators $X$, $Y$, $Z$, while $\rho_i$ is the density matrix of the corresponding eigenvectors with eigenvalues $c_i = \pm 1$. The expression Eq.~\eqref{eq:rho} can be simulated by a quasiprobability sampling (QPS) \cite{piveteau2022quasiprobability,pashayan2015estimating} (due to the sign of $c_i$) of two sub-circuits. In the measurement circuit  the expectation value of $O_i$ is measured, corresponding to $\operatorname{Tr} (\rho\, O_i)$, where $i$ is the cut qubit index; In the preparation circuit the eigenstate of $O_i$ is prepared in accordance with $\rho_i$. The results from the two sub-circuits are then combined to obtain an estimate for the expectation value of the corresponding operator. Due to QPS, circuit cutting generally incurs an exponential sampling overhead since an increased number of samples are required to represent the entanglement in the cut section of the circuit. This factor is determined by the type and number of gate/wire cuts, whether classical communication between the circuit partitions is allowed at execution time, and the specific decomposition used for the gates/wires that are being cut \cite{brenner2023optimal,harada2024doubly,ibm}. Depending on the choice of circuit cutting method, a small number of ancilla qubits may have to be introduced to individual circuit partitions.

The measurement overhead of the Pauli circuit cutting technique scales as $O(8^{k})$, where the base $8$ is related to the number of eigenstates of the operators and $k$ is the number of cuts in the circuit. To reduce this overhead, it is possible to only perform measurements in the $X$ and $Y$ bases in addition to the computational basis. Therefore, the preparation circuit can be initialized with only four specific states: $|0\rangle$, $|1\rangle$, $|+\rangle$, and $|i\rangle$. From these initial states, the expectation values for the $|-\rangle$, and $|-i\rangle$ states can be inferred using classical post-processing. The overhead for this circuit cutting method can thus be reduced to $O(4^k)$ per circuit.

In the context of sampling, circuit cutting has so far only been explored within the QAOA algorithm for Max-Cut and vehicle routing problems~\cite{bechtold2023investigating,ufrecht2024optimal}. It has been shown to help in reducing the noise introduced during the execution of such circuits. Although a spread in the probability distribution of the sampled states may occur for small circuit instances, the independence of the required number of measurements for cut circuits on the system size suggests that this effect should diminish as the circuit size is increased. In this work, we investigate the impact of circuit cutting on the quality of the basis generated within QSCI, in particular, for the computation of the DOS.


\begin{figure}
  \begin{subfigure}{0.48\textwidth}
    \includegraphics[width=\linewidth]{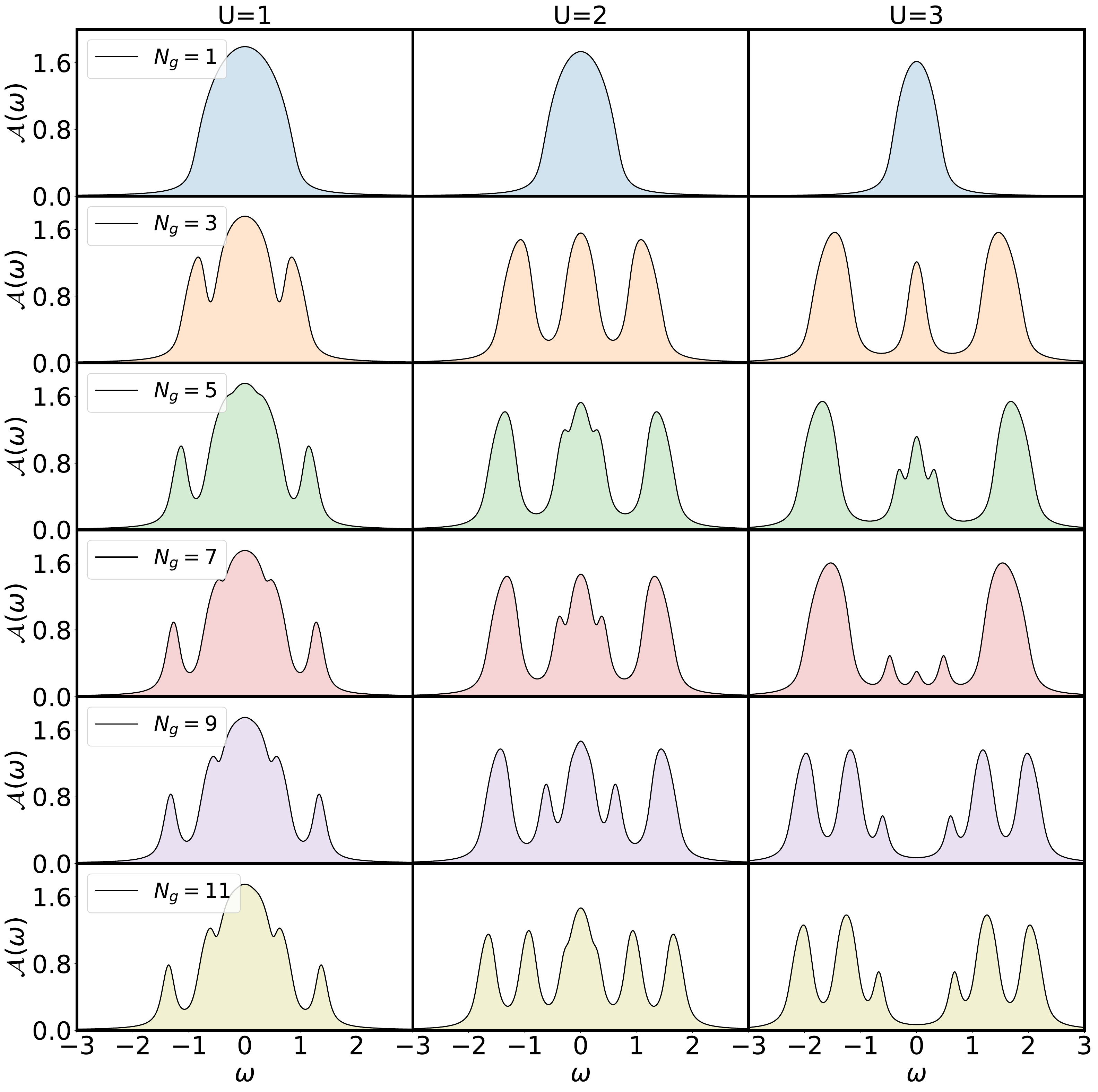}
    \vspace{-2\baselineskip}
    \caption{} \label{fig:2a}
  \end{subfigure}%
  \vspace{0.001cm}
  \begin{subfigure}{0.48\textwidth}
    \includegraphics[width=\linewidth]{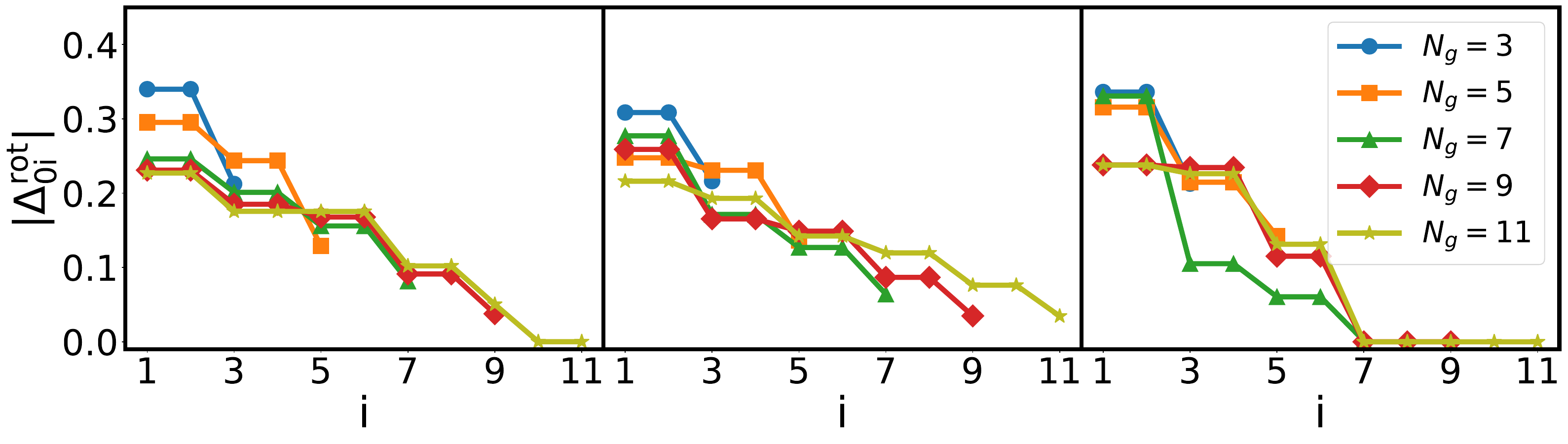}
    \vspace{-2\baselineskip}
    \caption{} \label{fig:2b}
  \end{subfigure}%
  \vspace{-1\baselineskip}
\caption{\justifying (a) Classical gGut calculations for the single-band Fermi-Hubbard model on the Bethe lattice using FCI as the impurity solver. The density of states $\mathcal{A}(\omega)$ is plotted for different numbers of ghosts $N_g = \{1,3,5,7,9,11\}$ and interaction strengths $U=\{1,2,3\}$. (b) The magnitude of the effective interactions between the impurity and the ghost orbitals $|\Delta_{0i}^{\text{rot}}|$, as obtained from the converged gGut calculations of panel a), where the ghosts are indexed by $i \in {1 \dots N_g}$ in descending order of their magnitude. The corresponding embedding Hamiltonians have been rotated into the star configuration (see main text).} \label{fig:dosclassical}
\end{figure}

\section{Results and Discussion}\label{sec:results}

\subsection{SCI impurity solver for gGut}
\label{sec:classical}

For the remainder of this work we focus on the Hubbard model on the Bethe lattice, a prototypical system used to benchmark embedding methods. At $U=0$, the non-interacting density of states, $\mathcal{A}_0(\omega)$, is semicircular. When a non-zero interaction strength $U$ is introduced, the quasi-particle peak around the Fermi level ($\omega = 0$), of the density of states gradually narrows and its height increases. This is a signature of Fermi liquid behavior, where electrons can be represented by well-defined quasi-particles, whose effective mass is enhanced due to the presence of interactions. As $U$ increases further, the central peak begins to split and the spectral weight is gradually transferred to two side peaks away from $\omega = 0$ called Hubbard bands. At the critical value of $U\approx 2.9$, the spectral weight at $\omega=0$ drops to $0$, thus opening a gap that signifies a first-order phase transition from a metal to a Mott insulator\cite{sriluckshmy_2021}. 


These features are also reflected in Fig.~\ref{fig:dosclassical}a where, using gGut, we map out the DOS as a function of the interaction strength $U$ and the number of ghosts $N_g$. We investigate the model for up to $N_g=11$ ghosts, which constitutes the largest single-site impurity gGut calculation to date. For a single ghost, gGut is equivalent to the Gutzwiller ansatz (GA) algorithm and, as reported in previous studies~\cite{lee2023accuracy_single}, fails to capture the metal-to-insulator phase transition altogether, instead producing a metallic DOS for all values of $U$. On the other hand, for small bath sizes where $1 < N_g \le 7$, we observe a qualitative change in the nature of the DOS, as the single peak slowly splits and Hubbard bands appear with increasing interaction strength. However, the transition occurs at values of $U > 3$, implying that gGut results have not yet fully converged in the number of ghosts. As the number of ghosts increases further ($N_g \geq 9$), we see that features in the DOS start to stabilize as a function of $N_g$. In particular, the shapes of the DOS for $N_g = 9$ and $N_g=11$ at $U=1$ are virtually the same, and a clear gap is visible for both bath sizes at $U=3$. In contrast, additional ghosts are required to stabilize the features of the DOS for $U=2$.  

\begin{figure}
\centering
\includegraphics[width=0.5\textwidth]{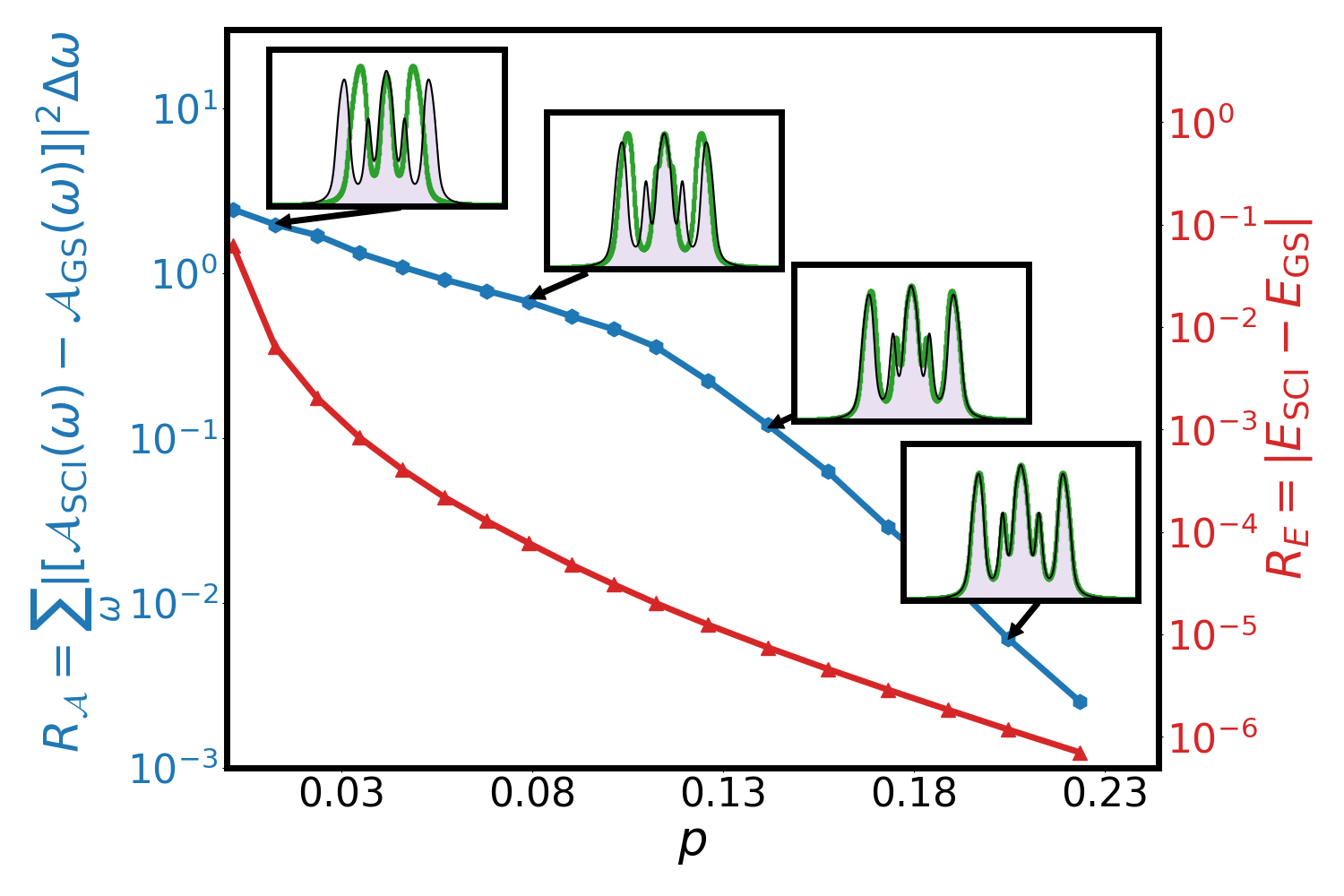}
\vspace{-2\baselineskip}
\caption{\justifying SCI impurity solver for gGut.  A comparison of the error in the DOS ($R_{\mathcal{A}}$, left axis, blue) and the energy ($R_{E}$, right axis, red) as a function of the truncation in the number of CIs selected by magnitude from the FCI ground state for $U=2$ and $N_g=9$. Here, $p$ represents the fraction of included basis states from $|\mathcal{S}_{\text{Sym}}|$. Insets show the progression of the DOS as a function of an increasing number of CIs included in the SCI set.}
\label{fig:errorsinenergf}
\end{figure}

\begin{figure*}[ht!]
\centering
\includegraphics[width=0.8\textwidth]{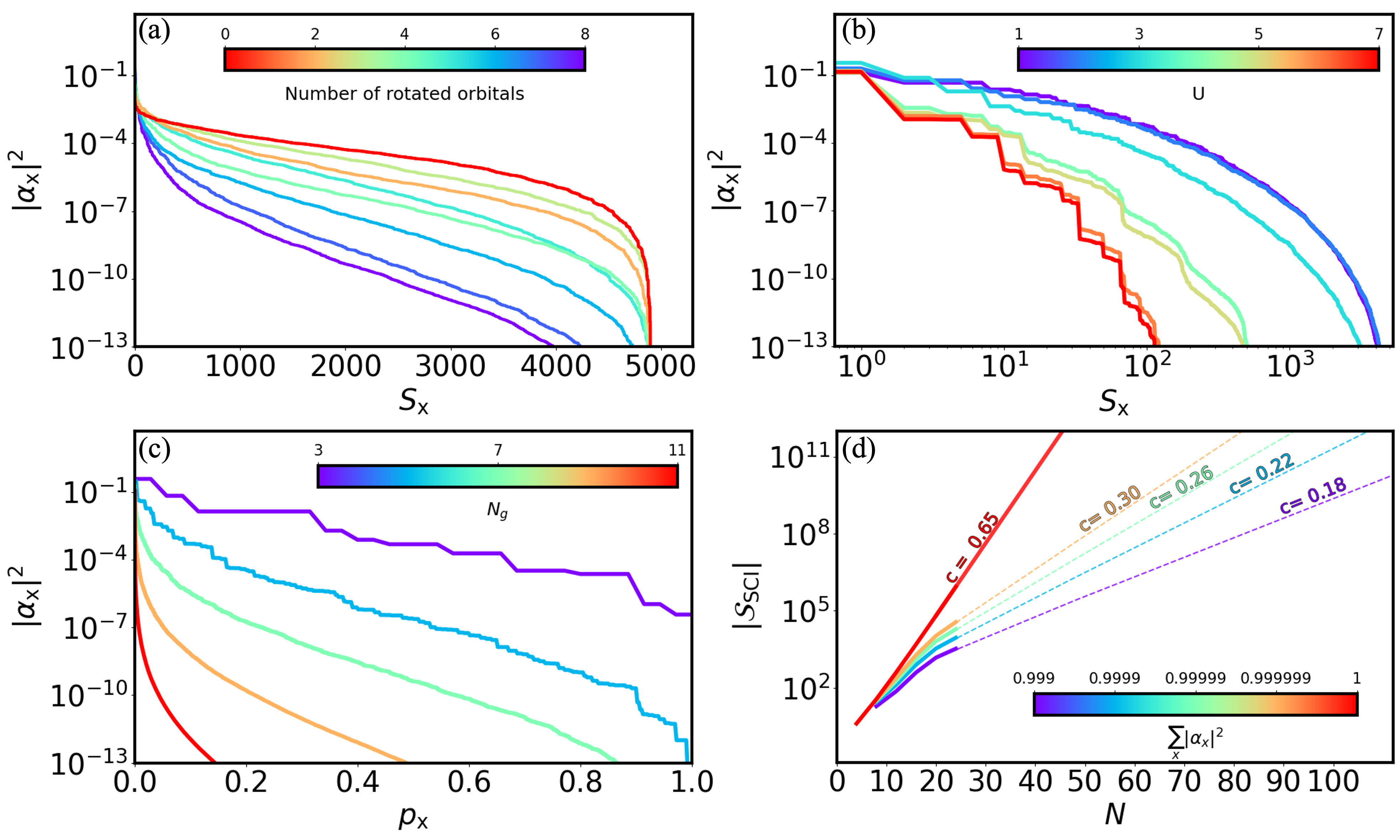}
  \vspace{-0.2cm}
\caption{\justifying (a) A comparison of the weights $|\alpha|^2$ of the CIs $S_x$ from the  ground state wavefunction, ordered by their magnitude, at $U=2$ and $N_g = 7$ is shown for different Hamiltonian bases, obtained by rotating the non-interacting part of $\hat{\mathcal{H}}^{\text{emb}}$ (see main text). (b) For the star configuration - the rotation indexed $7$, a comparison of the weights of CIs in the ground state wavefunction is shown for different interaction strengths $1\leq U \leq 7$ at $N_g=7$. (c) CI weights are shown for different bath sizes $N_g$ at $U=2$ and using the star configuration. The x-axis is rescaled for each bath size by the total number of CIs, $p_x = S_x/|\mathcal{S}_{\text{sym}}(N_g)|$ the values for which are tabulated in Appendix~\ref{appendix:details}. (d) The scaling of the  set $|\mathcal{S}_{\text{SCI}}|$, required to reproduce a fixed degree of accuracy of the ground-state wavefunction is shown as a function the total number of orbitals and extrapolated system sizes beyond $N=24$ using the fit $a+b\exp(c \,N)$. The total number of CIs in 
$|\mathcal{S}_{\text{sym}}|$ is shown for comparison (red).}
\label{fig:weightscomp}
\end{figure*}
 Fig.~\ref{fig:dosclassical}b shows the values of the converged rotated hybridization function acting between the impurity and the ghost orbitals, $\Delta^{\text{rot}}_{0i}$, sorted by magnitude after rotation to the star configuration, and as a function of $N_g$ and $U$. These values further corroborate the DOS results of Fig.~\ref{fig:dosclassical}a, as for $U=1$ and $U=3$, there is little change beyond $N_g=9$ and, in fact, hybridization terms with higher indices $i\geq 9$ and $i\geq 7$, respectively, are close to zero. For $U=2$, as before, the values for $\Delta^{\text{rot}}$ have not yet converged in the number of ghosts. 
 
We have observed convergence issues and result fluctuations within the gGut loop at larger system sizes, especially for $N_g=11$. These can be treated to a degree by introducing the mixing of results for multiple loop iterations which may, however, slow down the rate of convergence. Previously converged results from different $U$ values can also be used as a starting point to reduce the number of iterations and stabilize the loop, and $U$ can even be varied as a function of the loop iteration. This might still lead to slow convergence around phase transitions due to a qualitative difference between solutions on either side of the transition. Calculations beyond $N_g=11$ would thus require a switch from an FCI impurity solver to some other computationally more efficient method, such as the DMRG solver \cite{white1992density} used for multi-band Hamiltonians in Ref.~\cite{zhai2023block2,lee2023accuracy_multi}.

SCI could potentially serve as another efficient albeit cheap impurity solver alternative, but the algorithm has not yet been explored in the context of gGut. The main question in that respect is how sparse the ground state is in the CI basis, or, in other words, what fraction of the CI basis states is required to adequately capture the behavior of the system. To this end, we study the errors involved in the calculation of the ground state energy $E_{\text{GS}}$ and the corresponding density of states $\mathcal{A}(\omega)$ as a function of the size of the truncated SCI basis $|\mathcal{S}_{\text{SCI}}|$. Here, we build the $\mathcal{S}_{\text{SCI}}$ set from a fixed number of basis states ordered by the magnitude of their weights in the ground state wavefunction. Fig.~\ref{fig:errorsinenergf} investigates, for $N_g = 9$ and $U=2$, the error in the ground state energy, $R_E$, and the error in the DOS, $R_{\mathcal{A}}$, which are computed using:
\begin{align}
    R_{\mathcal{A}} & = \sum_\omega |\mathcal{A}_{\text{SCI}} (\omega) - \mathcal{A}_{\text{GS}} (\omega) |^2 \Delta \omega\\
    R_E & = |E_{\text{SCI}} - E_{\text{GS}}|.
\end{align}
Fig.~\ref{fig:errorsinenergf} demonstrates that $R_E$ converges considerably faster than $R_{\mathcal{A}}$, which is especially pronounced for a low number of CI basis states with $p \le 0.1$, where $p$ is the fraction of CIs from $\mathcal{S}_{\text{sym}}$ included in $\mathcal{S}_{\text{SCI}}$, and the total number of states being  $|\mathcal{S}_{\text{sym}}| = 63504$. Information about the resolution of the dynamical structure of the DOS can also be extracted from the investigation of its the higher moments~\cite{sriluckshmy_2021}. In Appendix~\ref{appendix:moments}, we compare the dependence of the error in higher moments of the DOS on the number of CIs in the SCI basis set. We find that in order to capture the DOS behavior at the level of higher moments, only a small constant overhead is required in terms of the the size of the CI basis. We observe that in order to correctly recover the features of the DOS, at least $|\mathcal{S}_{\text{SCI}}| \sim 10000$ CI's are required, roughly corresponding to $p=0.2$ and note that for this number of CI states one obtains a relative error  on the ground state energy below $10^{-5}$. We have also observed similar trends for other interaction strengths $U$ and numbers of ghosts $N_g$, albeit the specific values may vary, i.e. a lower CI fraction is required for larger $N_g$ and away from the phase transition. 

Next, we consider the influence of the single-particle basis choice on the convergence of SCI.
The ordered weights $\alpha_x$ of CIs $S_x \in \mathcal{S}_{\text{sym}}$ are plotted in Fig.~\ref{fig:weightscomp}a for $N_g=7$ and $U=2$, and as a function of the number of rotated orbitals. Here, the original basis obtained through gGut is indexed $0$, while the chain basis corresponds to index $1$ since a single rotated ghost orbital in the chain/tri-diagonal basis leaves the Hamiltonian unchanged. Indices $2-7$ correspond to the number of rotated ghost orbitals, with $7$ being the star configuration. Finally, we index the canonical basis with $8$. From Fig.~\ref{fig:weightscomp}a we see that CI weights drop the fastest in the canonical basis \cite{lu2019natural}. However, this sparsity comes at the cost of an impractical overhead from having to evaluate $O(N^4)$ two-particle Hamiltonian terms to construct the matrix corresponding to the rotated version of $\hat{\mathcal{H}}^{\text{emb}}$. One could, in principle, truncate these Hamiltonian terms to single and double excitation operators, which would reduce the aforementioned overhead at the price of an approximation. This, however, may prove detrimental to the computation of dynamical correlation functions, where, as we have learned from Fig.~\ref{fig:errorsinenergf}, a high accuracy of the wavefunction is required. The best trade-off between computational efficiency and a sufficiently fast decay rate of the CI weights is achieved for index $7$, the star configuration, where the ghost orbitals do not interact with each other, but individually couple to the impurity. In this case, the two-particle Hamiltonian terms in $\hat{\mathcal{H}}^{\text{emb}}$ remain unchanged, and any significant computational overhead can be avoided. This finding is in agreement with similar investigations in the context of tensor network algorithms \cite{kohn2021efficient}, where the entanglement of the star configuration was found to be significantly lower than the chain configuration, which was understood to be a result of the mixing of fully-occupied and empty conduction orbitals in the chain configuration. 

 \begin{figure*}
\centering
\includegraphics[width= 0.9\textwidth]{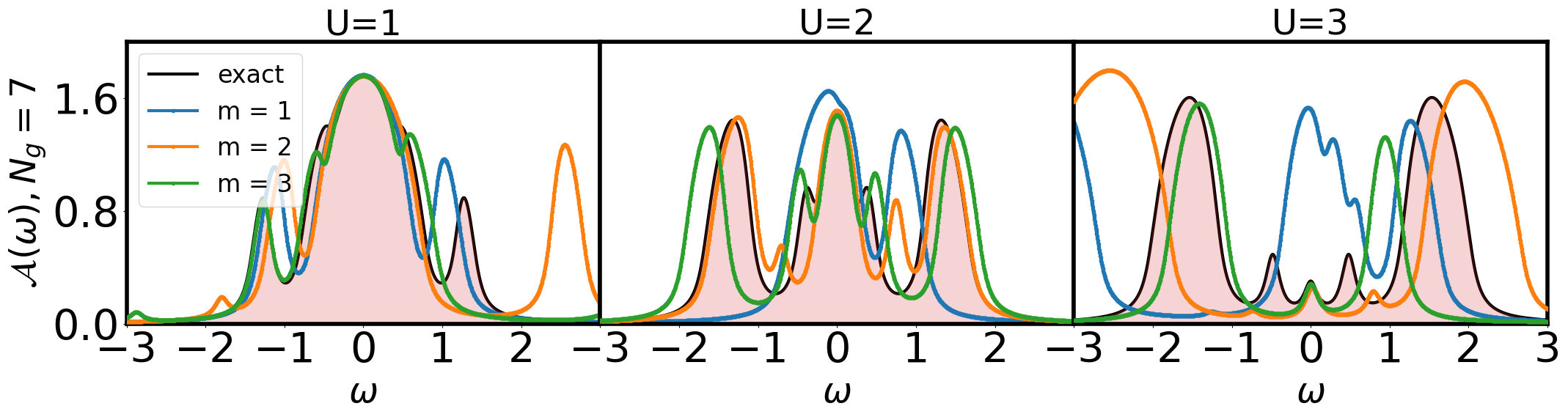}
\caption{\justifying Comparison of the approximate ground state DOS obtained from classically optimized and simulated LUCJ ansatz circuits with different numbers of ansatz layers, $m=\{1,2,3\}$, for interaction strengths $U=\{1,2,3\}$ and bath size $N_g=7$. Results are benchmarked against the exact DOS obtained from FCI (filled out in color).}
\label{fig:gf_lucj_ansatz}
\end{figure*}

Let us now focus on Fig.~\ref{fig:weightscomp}b which shows the dependence of the CI weights on the interaction strength $U$. Here, we consider $N_g=7$ and rotate the Hamiltonian into the star configuration. We observe that, as the interaction strength $U$ increases, the decay of the CI weights accelerates significantly. Consequently, fewer CIs are required to faithfully represent the ground state wavefunction in the strongly interacting limit. This reduced complexity can be partially understood from the fact that the Hilbert space in the insulating strong-coupling limit, $U\rightarrow\infty$, reduces to half of its size, where only a single fermion is allowed to occupy a pair of spin impurity orbitals. We found that the $U$-dependence of the ground state sparsity holds also for other choices of the single-particle Hamiltonian rotation and present similar findings for the rotation indexed $5$ in Appendix~\ref{appendix:udependence}. We also note that the sparsity will ultimately depend on the parameters of the converged embedding Hamiltonian $\hat{\mathcal{H}}^{\text{emb}}$ as obtained from the self-consistent gGut loop. Other embedding methods based on the Anderson Impurity model, such as DMFT\cite{georges1996dynamical} and EwDMET\cite{sriluckshmy_2021}, may thus result in ground states with a very different degree of sparseness in the CI basis. To further stress this point, we provide in Appendix~\ref{appendix:rotation} a  comparison of the CI weight decay for $\hat{\mathcal{H}}^{\text{emb}}$ obtained via gGut to a generic model with manually set values for the hybridization function $\Delta$, which match those used in Ref.~\cite{yu2025sample}.

Another central question is whether the SCI approach remains scalable as the number of ghosts $N_g$, and therefore the system size, is increased. To address this, we plot the CI weights $\alpha_{x}$ for different bath sizes and as a function of the normalized CI index $p_{x} = S_x/|\mathcal{S}_{\text{sym}}(N_g)|$ in Fig.~\ref{fig:weightscomp}c. We observe that the fraction of CIs above a given threshold weight decreases with system size, meaning that the ground state sparsity increases, which further supports the findings of Fig.~\ref{fig:dosclassical}b. 

In Fig.~\ref{fig:weightscomp}d, we study the rate of increase in the number of CIs required to reach a certain ground state fidelity, $\Sigma_\alpha = \sum_x |\alpha_x|^2$,  as a function of $N$ ($2(N_o + N_g)$) and compare it to the rate of growth of $|\mathcal{S}_{\text{sym}}|$. From Fig.~\ref{fig:errorsinenergf}, we can derive that, at least for $N_g=9$ (N = 20), we can capture the correct qualitative behavior of the DOS with $\Sigma_\alpha=0.9999$ and recover the exact quantitative values of the DOS for all frequencies with $\Sigma_\alpha=0.999999$. From fitting data for up to $N_g=11$ (equivalent to $N=24$) with the function $a+b \exp(c\,N)$, we find that the number of required CIs increases exponentially, although with a lower coefficient compared to $\mathcal{S}_{\text{sym}}$, and depends on the choice $\Sigma_\alpha$. Specifically, we get $c=0.22$ for $\Sigma_\alpha=0.9999$ and $c=0.30$ for $\Sigma_\alpha=0.999999$. Extrapolating using these exponential fits, we find that if we allow up to $10^{12}$ CIs in the SCI basis set, which corresponds to the number of CIs in current state-of-the-art numerical implementations of FCI~\cite{gao2024distributed}, we can reach $N_g=41$ and $N_g=53$, for $\Sigma_\alpha=0.999999$ and $\Sigma_\alpha=0.9999$,
 respectively. This corresponds to $84-108$ qubits on a quantum computer assuming that the Jordan-Wigner transformation is used. In comparison, a full FCI calculation based on the set $\mathcal{S}_{\text{sym}}$ reaches the same CI limit at $N_g=23$, which is equivalent to $48$ qubits. We wish to emphasize that the reach of SCI may actually be even higher, since the data in Fig.~\ref{fig:weightscomp}d does not look to be fully converged and, in fact, a deviation is observed at $N_g=11$.

\begin{figure*}
\centering
\includegraphics[width=0.9\textwidth]{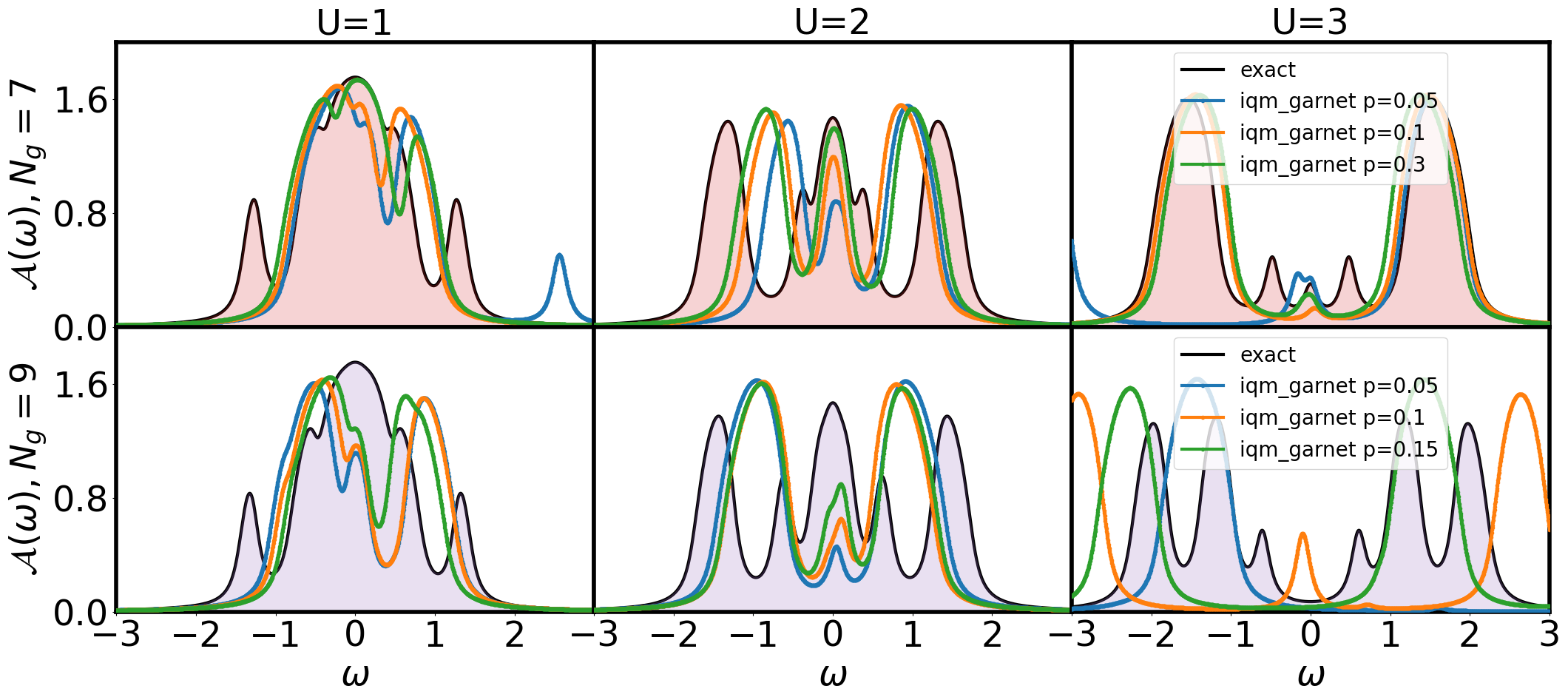}
\caption{\justifying Comparison of the DOS computed with QSCI constructed with $3.2 \cdot 10^5$ samples obtained from an \emph{uncut} LUCJ ansatz circuit with $m=1$ layers, and prepared on IQM Garnet quantum hardware. The QSCI basis is truncated at $p=\{0.05,0.1,0.3\}$ for $N_g=7$ and at $p=\{0.05,0.1,0.15\}$ for $N_g = 9$. Results are benchmarked against the exact DOS obtained from FCI (filled out in color) for values of the interaction strength of $U=\{1,2,3\}$.}
\label{fig:nocut}
\end{figure*}

\subsection{QSCI impurity solver for gGut}
\label{sec:quantum}

Motivated by the classical SCI results of the previous section, we now focus our attention on implementing QSCI as an impurity solver for gGut. In particular, we have so far assumed that CIs can be sampled from the FCI ground state wavefunction, which would in practice make the SCI computation redundant. Whilst numerous classical heuristics exist for this purpose, we are interested in exploring the sampling from quantum trial states to build the SCI basis. In what follows, the quantum trial states are prepared using the LUCJ ansatz as described in Section~\ref{sec:lucj}. 

First, we study the potential of such parameterized quantum trial states to directly capture the correct behavior of the DOS. Since LUCJ ansatz states for up to $N_g=11$, corresponding to 24 qubits, can be efficiently prepared on classical computers, we compute the noiseless approximate ground states with classically pre-optimized circuit parameters. Here, the optimization procedure is warm-started from a coupled cluster singles-doubles (CCSD) \cite{bartlett2007coupled} calculation in order to reduce the number of iterations. For this optimization, we utilize the already converged Hamiltonian parameters of $\hat{\mathcal{H}}_{\text{emb}}$ from the self-consistent gGut loop. In Fig.~\ref{fig:gf_lucj_ansatz} we plot the DOS obtained from LUCJ states with up to $m=3$ ansatz layers for $U=\{1,2,3\}$ and $N_g=7$, and benchmark them against the exact DOS obtained from FCI (see Fig.~\ref{fig:dosclassical}). While the results for the DOS improve with an increasing number of LUCJ layers, even the $m=3$ state is insufficient to accurately capture the functional form of the DOS away from the low-frequency regime. At higher frequencies, on the other hand, we observe the appearance of spurious peaks in the DOS. This shortcoming also persists for higher values of $N_g$, as we show in Appendix \ref{appendix:lucjmore}, and in fact, becomes worse since the ansatz fails to even qualitatively capture the gap in the DOS at zero frequency (see Fig.~\ref{figapp:gf_lucj_ansatz_add} of the Appendix). We expect that, due to quantum hardware noise, these results would further deteriorate if they were performed on real quantum devices, e.g. using the variational quantum eigensolver (VQE) algorithm \cite{peruzzo2014variational}. 

Next, the LUCJ ansatz is prepared on the quantum hardware wherein for an implementation of $\tilde{K}_{m}$, $N/2(N/2-1)$ entangling gates of the  $\exp(i \theta_{ij}^{m\sigma} (X_i^\sigma X_j^\sigma+Y_i^\sigma Y_j^\sigma))$ type together with $N$ single-qubit rotation gates $R_z$ are required, resulting in a total circuit depth of $N/2+1$. The Jastrow terms $\prod_{\langle i,j\rangle,\sigma} \tilde{J}_{m}$, on the other hand, require only $N-2$ gates of the $\exp (i \phi_{ij}^{m\sigma} Z_i^{\sigma} Z_j^{\sigma})$ type which can be implemented in a constant depth of $2$~\cite{motta2023bridging}. Finally the $\tilde{J}_{m}^{\uparrow \downarrow} $ only involves a single $\exp(i \phi_{ij}^{m\uparrow\downarrow} Z_0^{\uparrow} Z_0^{\downarrow})$ gate which can be executed in parallel with the other Jastrow terms. The quantum circuit for for $N=20$ qubits ($N_g=9$) transpiles into a total of $407$ native CZ gates on the IQM quantum hardware, corresponding to a circuit depth of $234$. 

\begin{figure*}
\centering
\includegraphics[width=0.9\textwidth]{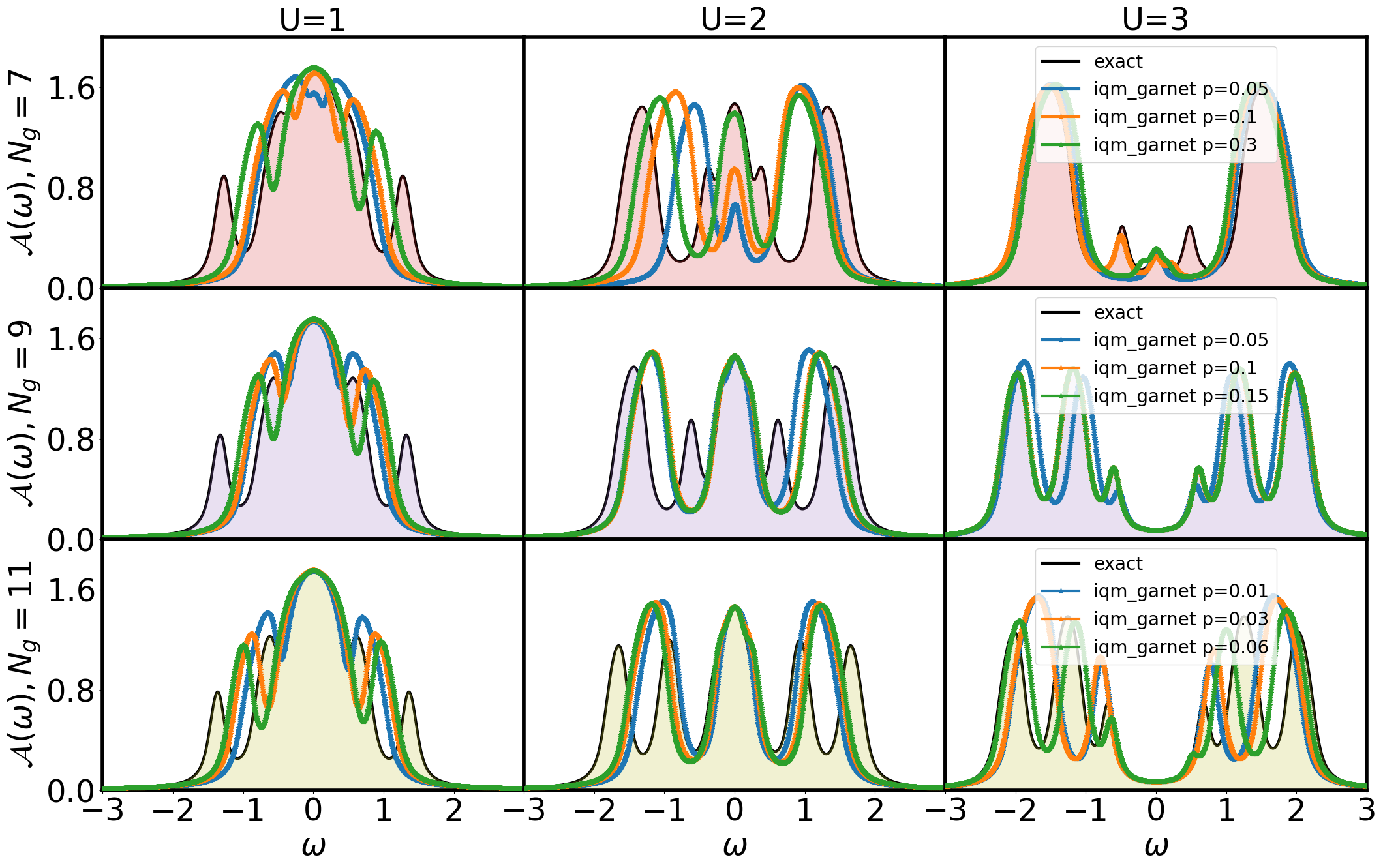}
\caption{\justifying Comparison of the DOS computed with QSCI plus circuit cutting, constructed with a total $3.2 \cdot 10^5$ samples ($2 \cdot 10^4$ per sub-circuit) obtained from an LUCJ ansatz circuit with $m=1$ layers where two wire cuts have been performed, and prepared on IQM Garnet quantum hardware. Results are shows as a function of the interaction strength $U=\{1,2,3\}$, the bath size $N_g = \{7,9,11\}$ and the QSCI basis truncation fraction  $p$. Results are benchmarked against the exact DOS obtained from FCI (filled out in color).}\label{fig:circcut}
\end{figure*}

\begin{figure*}
\centering
\includegraphics[width=0.9\textwidth]{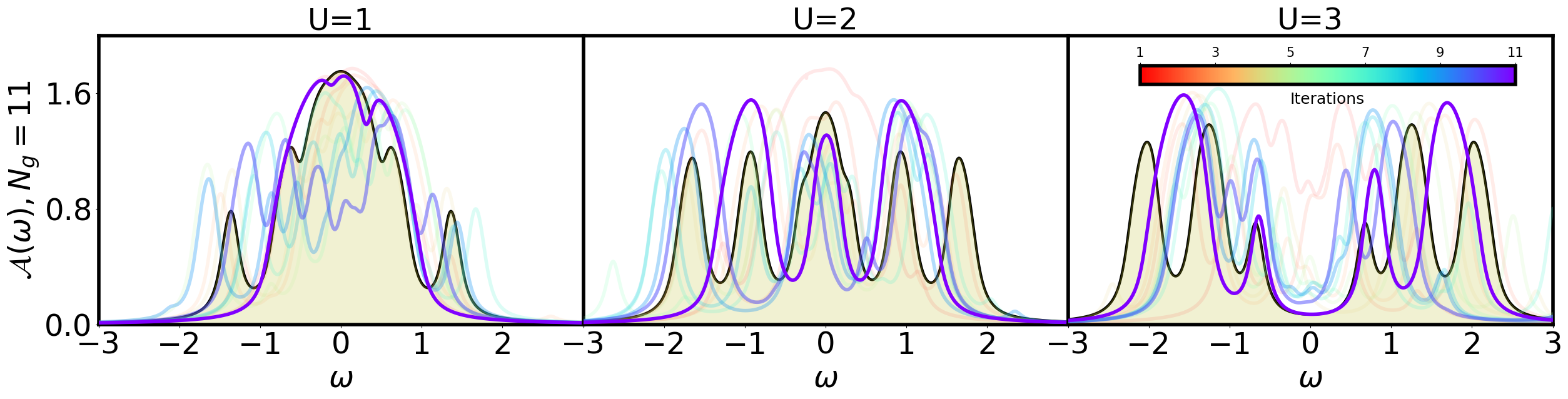}
\caption{\justifying Comparison of the DOS obtained with QSCI plus circuit cutting, computed from $11$ iterations of the self-consistent gGut loop, each with different parameters for $\hat{\mathcal{H}}^{\text{emb}}$. At each loop iteration, the SCI basis was constructed from a total of $3.2 \cdot 10^5$ samples truncated to $p=0.01$. An LUCJ ansatz circuit with $m=1$ layers and two wire cuts, prepared on IQM Garnet quantum hardware was used as trial state. The DOS is shown as a function of the iteration for interaction strengths $U=\{1,2,3\}$ and $N_g=11$. Results are benchmarked against the exact DOS obtained from FCI (filled out in color).}\label{fig:fullgga}
\end{figure*}

We investigate a quantum implementation of QSCI, which we execute on the IQM Garnet 20-qubit quantum device \cite{abdurakhimov2024technology}.
Fig.~\ref{fig:nocut} shows our QSCI results for the DOS with $U=\{1,2,3\}$ and $N_g=\{7,9\}$. We generate the set $\mathcal{S}_{\text{SCI}}$ from $3.2\cdot10^5$ samples of a pre-optimized LUCJ circuit with $m=1$, equivalent to the trial states used for Fig.~\ref{fig:gf_lucj_ansatz}. From these samples, we truncate the basis to various percentages of the total number of CIs of $\mathcal{S}_{\text{sym}}$ and pick the CIs based on their sampled weights. To reduce the impact of noise, we post-select the CIs within the correct (number and spin) symmetry sector (see Fig~\ref{fig:main}h). We have also attempted to improve results through the use of readout error mitigation \cite{Cai_2023_QEMreview}, but found that it had minimal impact on the quality of the CI basis, as is further detailed in Appendix \ref{appendix:rem}. For comparison we also plot the simulated results, including only sampling noise, which demonstrates that comparable features to the noisy hardware results can be achieved with even fewer CIs. From Fig.~\ref{fig:nocut} we see that, despite the use of a real quantum device, the DOS results from QSCI show an improvement over Fig.~\ref{fig:gf_lucj_ansatz}, especially in consistently describing the low-frequency behavior of the DOS across the interaction strength $U$ induced phase transition. This means that the SCI basis, sampled from the LUCJ state, manages to capture the most relevant CIs despite the presence of hardware noise. As the number of CIs in the set $\mathcal{S}_{\text{SCI}}$ increases, there is an improvement in the distribution of the spectral weight, with a shift observed from spurious high-frequency peaks towards the central low-frequency region. However, these results are still quantitatively lacking and appear far from converged even at $p=0.3$ and $p=0.15$, for $N_g=7$ and $N_g=9$, respectively. 

As an alternative implementation of QSCI, we use circuit cutting to decompose the same LUCJ ($m=1$) ansatz circuit, as described in Sec.~\ref{sec:circuitcutting} (see also Fig~\ref{fig:main}g). Specifically, two wire cuts are performed before and after the density-density interaction gate ($\gamma_{\uparrow\downarrow}$) between the two qubits assigned to impurity orbitals, in order to decouple the circuit into spin-up and spin-down parts, resulting in a manageable $16\times$ overhead in the number of executed circuits and one ancilla qubit per sub-circuit. For for $N=24$ ($N_g=11$), the total number of transpiled native CZ entangling gates on the IQM quantum hardware per sub-circuit is $286-290$, with a corresponding circuit depth of $142-145$. We used $2\cdot10^4$ shots for each of the cut circuits, which amounts to the same $3.2\cdot10^5$ samples as have been used for the uncut circuit. The depth of the sub-circuits remain the same as the uncut circuit whereas the number of gates is reduced by a factor of two. The gates, however, can become non-local due to the ancilla qubit introduced by the cutting. We stress that, when circuit cutting is used, the CIs must be reassembled from the sampled partial CIs for each spin through a tensor product. As a consequence, this leads to different CIs as well as different CI weights within $\mathcal{S}_{\text{SCI}}$, compared to the \emph{uncut} QSCI procedure. As shown in Fig.~\ref{fig:circcut} for $N_g=\{7,9,11\}$, there is a significant overall improvement in the resulting DOS compared to the \emph{uncut} QSCI. In particular, the low frequency peaks at small $U$ as well as the opening of the gap at higher $U$ are much better captured at similar truncation levels to Fig.~\ref{fig:nocut}. The results for $N_g = 11$ show that with as low as $1\%$ of the total CI states, the gap in the DOS is well resolved, and with a truncation at $p=0.06$ even some of the side bands at high frequency can be captured for $U=3$. A higher number of layers for the LUCJ ansatz ($m>1$) could be used to further improve these results, which would require an increased sampling overhead due to the necessity of performing $2m$ wire cuts.

To test the robustness of the combined QSCI plus circuit cutting technique introduced above, we now aim to use it as an impurity solver inside of the self-consistent gGut loop (as described in step 3 of Sec.~\ref{sec:method}). At each step of the loop, a new embedding Hamiltonian $\hat{\mathcal{H}}^{\text{emb}}$ is constructed from $\Delta$ and $\Lambda^{\text{emb}}$, and a single-particle basis rotation to the star geometry is performed. We use the single layer ($m=1$) LUCJ ansatz, the parameters of which are determined using $5$ iterations of the simultaneous perturbation stochastic approximation (SPSA)\cite{spall1998overview} optimization procedure executed on a noiseless simulator for the ansatz circuit. Circuit cutting is applied to the ansatz circuit and from the measurements of the sub-circuits on classical computers, the CI basis for QSCI is reassembled and truncated at $1\%$ of the total number of basis states in $\mathcal{S}_{\text{sym}}$. Subsequently, the sub-circuits corresponding to the SPSA optimized ansatz are executed on IQM Garnet and QSCI performed with the same truncation. From the approximate ground state wavefunction $\Psi_{\text{SCI}}^{\text{emb}}$, the 1-RDMs $\rho^{\text{emb}}$ and $\zeta$ are computed, and these are, in turn, injected into step $4$ of the gGut loop. Our results for the DOS are shown in Fig.~\ref{fig:fullgga}. Specifically, we set the bath size to $N_g = 11$ and study interaction strengths $U=\{1,2,3\}$. The DOS is shown as a function of the loop iteration, with $11$ iterations performed in total. We notice that the gGut loop is stable under the use of this hybrid quantum-classical impurity solver even for this largest of all studied system sizes. 
The qualitative behavior of the DOS in the low frequency regime remains very well captured and  the gap is accurately reproduced for $U=3$. Similar to our findings from Fig.~\ref{fig:circcut}, a higher percentage of CIs may be required for the solver to also resolve the side bands at higher frequencies. Nevertheless, the stability and convergence of the self-consistent gGut loop are encouraging results for the potential future use of QSCI as an impurity solver in quantum embedding methods. 

\section{Conclusions}\label{sec:conclusions}

In this work, we have studied the applicability of SCI-based techniques as impurity solvers within the gGut embedding algorithm. We have tested our approaches on the single-band Fermi-Hubbard model defined on the Bethe lattice across the metal-to-insulator phase transition as a function of U, which can be identified from the frequency dependence of the density of states. We have successfully converged the gGut self-consistency loop for embedding Hamiltonians with bath sizes of up to $N_g=11$. Beyond this bath size, we found the convergence of  gGut prohibitively challenging, particularly due to emerging stability issues of the self-consistent loop. While this may be an artifact of the investigated model, and in fact overall larger converged calculations have been reported for multi-band systems~\cite{lee2023accuracy_multi}, it will be paramount to implement tools that can reinforce algorithmic stability going forward. In this context, it would be insightful to compare the performance of SCI to alternative impurity solvers used in embedding techniques, such as the DMRG from Ref.~\cite{lee2023accuracy_multi}.

We have shown that, using an appropriate single-particle Hamiltonian basis rotation, the ground state of this model is sufficiently sparse, thus allowing for SCI to significantly extend the reach of FCI, potentially facilitating calculations for systems with more than 100 fermionic modes. It would be of great interest to extend this study to gGut implementations for other models, such as multi-band periodic solids obtained from ab-initio DFT or GW calculations \cite{lee2023accuracy_multi}, molecular Hamiltonians \cite{mejuto2024quantum}, and non-equilibrium systems \cite{guerci2023time}, which will be addressed in future studies. 

We found that the sparsity of the ground state depends not only on system parameters such as $U$, but also on whether model parameters, such as $\Delta$ are taken ad hoc or from converged gGut calculations. Further, a much larger number of CIs is required to reach the same precision on dynamic quantities such as the density of states compared to static quantities such as the energy. The smaller the bath size, the larger the percentage of CIs which is required to reach a given precision on either of these quantities. From classical runs with LUCJ, we observed that it correctly captures the hierarchy of most relevant CIs when compared to FCI.

Moving beyond classical gGut computations, we have investigated QSCI as an impurity solver and found an overall promising performance when the hybrid method was implemented on IQM's superconducting quantum hardware. In particular, we reproduced qualitatively, and in some cases even quantitatively, the correct behavior of the DOS across the metal-to-insulator phase transition and for bath sizes of up to $N_g=11$ ghost orbitals, corresponding to quantum circuits with up to 24 qubits. We showed that the low frequency features of the DOS can be reproduced with a low fraction of $1\%$ of the total CIs for $N_g=11$ ghosts and for all studied values of $U$. We note that, even noiseless LUCJ trial states without the SCI step, would yield severely inferior results in terms of reproducing the ground state properties of the system. 

Interestingly, we found that the use of circuit cutting had a significant positive impact on our QSCI calculations despite its computational overhead in the number of circuit executions. This improvement can be credited to reduced hardware noise in smaller circuits, but also to the mixing of CIs from partial circuits in the reassembly process, which effectively changed the nature of the underlying quantum states.  We note that classical communication may not necessarily improve the sampling overhead due to circuit cutting, which can instead be improved by cutting multiple gates simultaneously or cutting gates separated by black box operations \cite{schmitt2025cutting}. Parallel n-wire cuts could be further improved through the use of ancilla-free techniques based on mutually unbiased bases \cite{harada2024doubly}. Other, potentially more efficient circuit cutting schemes could also be explored~\cite{gentinetta2024overhead}.

To improve the quality of samples obtained from the quantum device, we have performed error mitigation in the form of post-selection based on symmetry verification and readout error mitigation. We found that while the former improved the QSCI performance, the latter did not have any significant impact on it. It would be worthwhile to develop and investigate alternative error mitigation methods tailored to sampling algorithms such as QSCI~\cite{liu2025quantum}. 
More generally, it is an open question which kind of quantum states perform best in the context of QSCI. In particular, such quantum states should ideally be out of reach of classical simulators, something that is not true for the LUCJ states at the system sizes we have implemented in this work. One recent promising direction includes the generation of a CI basis from real-\cite{mikkelsen2024quantum,yu2025sample} or imaginary-time evolution\cite{anuar2024operator, gluza2024double} of quantum states, which effectively approximates the Krylov basis and is not easily prepared on classical computers. As quantum computers become ever more sophisticated, the spectrum of possibilities to explore novel trial states is bound to grow.

To our knowledge, this work represents the first instance of a hybrid quantum-classical embedding algorithm of the quoted system sizes, where the self-consistent loop was shown to converge using data from actual quantum hardware. Despite this achievement, parameter optimization for the quantum state was still performed classically. For larger system sizes, such variational circuits are expected to become prohibitively hard to train, in particular due to the appearance of barren plateaus \cite{larocca2025barren}. However, we stress that it is sufficient for the quantum state to capture all relevant CIs, which may be easier to achieve compared to preparing a good approximation to the actual ground state, especially given recent improvements in classical simulation techniques~\cite{miller2025simulation}.

Finally, it would be of interest to study the applicability of the QSCI algorithm to other embedding methods, such as EwDMET and DMFT, which may produce embedding Hamiltonians with different structures and levels of sparseness. In the case of DMFT it would also require the computation of a number of excited states in order to correctly capture the single-particle Green's function, which is not investigated in this work, but 
is generally possible within the framework of QSCI \cite{kanno2023quantum}. Overall, this work has paved the way for further investigations into the use of (Q)SCI as an impurity solver in various embedding methods, with the promise of eventually extending the reach of simulations for real materials, including transition metals, rare-earth elements, and actinide compounds. 

\section*{Acknowledgments}
The authors thank Aeishah Ameera Binti Anuar, Alessio Calzona, Nicola Lanat\`a, Martin Leib and Aniket Rath for useful discussions. All FCI calculations for gGut have been performed using \textsc{QuSpin} \cite{weinberg2017quspin}. Benchmarking and pre-training of the LUCJ ansatz circuits has been done with \textsc{PySCF}~\cite{sun2020recent} and \textsc{ffsim}~\cite{ffsim} unless specified otherwise. 

After the completion of this manuscript, the authors came across a recently published manuscript \cite{chen2025quantum} that investigates the use of quantum imaginary time evolution (QITE) algorithm as an impurity solver within the gGut formalism to study the Fermi-Hubbard model on the Bethe lattice for up to $N_g=3$ ghosts. We note that QITE represents an alternative way of preparing quantum (trial) states compared to LUCJ circuits used here, which, in principle, could be used in combination with the QSCI plus circuit cutting technique presented in this work. We also note that, unlike this work, no self-consistent gGut calculation was performed in the aforementioned publication.

This work was supported by the German Federal Ministry of Education and Research (BMBF), within the Research Program Quantum Systems, via the joint project QUBE (grant number 13N17152). 

\bibliographystyle{unsrt}
\bibliography{refs}

\begin{thebibliography}{100}

\bibitem{troyer2005computational}
Matthias Troyer and Uwe-Jens Wiese.
\newblock Computational complexity and fundamental limitations to fermionic quantum monte carlo simulations.
\newblock {\em Physical review letters}, 94(17):170201, 2005.

\bibitem{georges1996dynamical}
Antoine Georges, Gabriel Kotliar, Werner Krauth, and Marcelo~J Rozenberg.
\newblock Dynamical mean-field theory of strongly correlated fermion systems and the limit of infinite dimensions.
\newblock {\em Reviews of modern physics}, 68(1):13, 1996.

\bibitem{maier2005quantum}
Thomas Maier, Mark Jarrell, Thomas Pruschke, and Matthias~H Hettler.
\newblock Quantum cluster theories.
\newblock {\em Reviews of Modern Physics}, 77(3):1027--1080, 2005.

\bibitem{yamazaki2018towards}
Takeshi Yamazaki, Shunji Matsuura, Ali Narimani, Anushervon Saidmuradov, and Arman Zaribafiyan.
\newblock Towards the practical application of near-term quantum computers in quantum chemistry simulations: A problem decomposition approach.
\newblock {\em arXiv preprint arXiv:1806.01305}, 2018.

\bibitem{jamet2021krylov}
Francois Jamet, Abhishek Agarwal, Carla Lupo, Dan~E Browne, Cedric Weber, and Ivan Rungger.
\newblock Krylov variational quantum algorithm for first principles materials simulations.
\newblock {\em arXiv preprint arXiv:2105.13298}, 2021.

\bibitem{ma2021quantum}
He~Ma, Nan Sheng, Marco Govoni, and Giulia Galli.
\newblock Quantum embedding theory for strongly correlated states in materials.
\newblock {\em Journal of Chemical Theory and Computation}, 17(4):2116--2125, 2021.

\bibitem{yao2021gutzwiller}
Yongxin Yao, Feng Zhang, Cai-Zhuang Wang, Kai-Ming Ho, and Peter~P Orth.
\newblock Gutzwiller hybrid quantum-classical computing approach for correlated materials.
\newblock {\em Physical Review Research}, 3(1):013184, 2021.

\bibitem{besserve2022unraveling}
Pauline Besserve and Thomas Ayral.
\newblock Unraveling correlated material properties with noisy quantum computers: Natural orbitalized variational quantum eigensolving of extended impurity models within a slave-boson approach.
\newblock {\em Physical Review B}, 105(11):115108, 2022.

\bibitem{greene2022modelling}
Gabriel Greene-Diniz, David~Zsolt Manrique, Wassil Sennane, Yann Magnin, Elvira Shishenina, Philippe Cordier, Philip Llewellyn, Michal Krompiec, Marko~J Ran{\v{c}}i{\'c}, and David~Mu{\~n}oz Ramo.
\newblock Modelling carbon capture on metal-organic frameworks with quantum computing.
\newblock {\em EPJ Quantum Technology}, 9(1):37, 2022.

\bibitem{iijima2023towards}
Naoki Iijima, Satoshi Imamura, Mikio Morita, Sho Takemori, Akihiko Kasagi, Yuhei Umeda, and Eiji Yoshida.
\newblock Towards accurate quantum chemical calculations on noisy quantum computers.
\newblock {\em arXiv preprint arXiv:2311.09634}, 2023.

\bibitem{dhawan2024quantum}
Diksha Dhawan, Dominika Zgid, and Mario Motta.
\newblock Quantum algorithm for imaginary-time green’s functions.
\newblock {\em Journal of Chemical Theory and Computation}, 20(11):4629--4638, 2024.

\bibitem{bertrand2024turning}
Corentin Bertrand, Pauline Besserve, Michel Ferrero, and Thomas Ayral.
\newblock Turning qubit noise into a blessing: Automatic state preparation and long-time dynamics for impurity models on quantum computers.
\newblock {\em arXiv preprint arXiv:2412.13711}, 2024.

\bibitem{shajan2024towards}
Akhil Shajan, Danil Kaliakin, Abhishek Mitra, Javier~Robledo Moreno, Zhen Li, Mario Motta, Caleb Johnson, Abdullah~Ash Saki, Susanta Das, Iskandar Sitdikov, et~al.
\newblock Towards quantum-centric simulations of extended molecules: sample-based quantum diagonalization enhanced with density matrix embedding theory.
\newblock {\em arXiv preprint arXiv:2411.09861}, 2024.

\bibitem{jamet2025anderson}
Fran{\c{c}}ois Jamet, Lachlan~P Lindoy, Yannic Rath, Connor Lenihan, Abhishek Agarwal, Enrico Fontana, Fedor Simkovic, Baptiste~Anselme Martin, and Ivan Rungger.
\newblock Anderson impurity solver integrating tensor network methods with quantum computing.
\newblock {\em APL Quantum}, 2(1), 2025.

\bibitem{ehrlich2025variational}
Jannis Ehrlich, Daniel~F Urban, and Christian Els{\"a}sser.
\newblock Variational quantum-algorithm based self-consistent calculations for the two-site dmft model on noisy quantum computing hardware.
\newblock {\em Journal of Physics: Condensed Matter}, 2025.

\bibitem{sheng2022green}
Nan Sheng, Christian Vorwerk, Marco Govoni, and Giulia Galli.
\newblock Green’s function formulation of quantum defect embedding theory.
\newblock {\em Journal of Chemical Theory and Computation}, 18(6):3512--3522, 2022.

\bibitem{rusakov2018self}
Alexander~A Rusakov, Sergei Iskakov, Lan~Nguyen Tran, and Dominika Zgid.
\newblock Self-energy embedding theory (seet) for periodic systems.
\newblock {\em Journal of chemical theory and computation}, 15(1):229--240, 2018.

\bibitem{lan2017generalized}
Tran~Nguyen Lan and Dominika Zgid.
\newblock Generalized self-energy embedding theory.
\newblock {\em The journal of physical chemistry letters}, 8(10):2200--2205, 2017.

\bibitem{tilly_2021}
Jules Tilly, P.~V. Sriluckshmy, Akashkumar Patel, Enrico Fontana, Ivan Rungger, Edward Grant, Robert Anderson, Jonathan Tennyson, and George~H. Booth.
\newblock Reduced density matrix sampling: Self-consistent embedding and multiscale electronic structure on current generation quantum computers.
\newblock {\em Phys. Rev. Res.}, 3:033230, Sep 2021.

\bibitem{sun2016quantum}
Qiming Sun and Garnet Kin-Lic Chan.
\newblock Quantum embedding theories.
\newblock {\em Accounts of chemical research}, 49(12):2705--2712, 2016.

\bibitem{vorwerk2022quantum}
Christian Vorwerk, Nan Sheng, Marco Govoni, Benchen Huang, and Giulia Galli.
\newblock Quantum embedding theories to simulate condensed systems on quantum computers.
\newblock {\em Nature Computational Science}, 2(7):424--432, 2022.

\bibitem{hettler2000dynamical}
Mattias~H Hettler, M~Mukherjee, Mark Jarrell, and Hulikal~R Krishnamurthy.
\newblock Dynamical cluster approximation: Nonlocal dynamics of correlated electron systems.
\newblock {\em Physical Review B}, 61(19):12739, 2000.

\bibitem{kotliar2001cellular}
Gabriel Kotliar, Sergej~Y Savrasov, Gunnar P{\'a}lsson, and Giulio Biroli.
\newblock Cellular dynamical mean field approach to strongly correlated systems.
\newblock {\em Physical review letters}, 87(18):186401, 2001.

\bibitem{rohringer2018diagrammatic}
G~Rohringer, H~Hafermann, A~Toschi, AA~Katanin, AE~Antipov, MI~Katsnelson, AI~Lichtenstein, AN~Rubtsov, and K~Held.
\newblock Diagrammatic routes to nonlocal correlations beyond dynamical mean field theory.
\newblock {\em Reviews of Modern Physics}, 90(2):025003, 2018.

\bibitem{klett2020real}
Marcel Klett, Nils Wentzell, Thomas Sch{\"a}fer, Fedor Simkovic~IV, Olivier Parcollet, Sabine Andergassen, and Philipp Hansmann.
\newblock Real-space cluster dynamical mean-field theory: Center-focused extrapolation on the one-and two particle-levels.
\newblock {\em Physical Review Research}, 2(3):033476, 2020.

\bibitem{pavarini2011lda}
Eva Pavarini, Dieter Vollhardt, Erik Koch, and Alexander Lichtenstein.
\newblock The lda+ dmft approach to strongly correlated materials.
\newblock Technical report, Theoretische Nanoelektronik, 2011.

\bibitem{held2001realistic}
Karsten Held, Igor~A Nekrasov, Nils Bl{\"u}mer, VI~Anisimov, and Dieter Vollhardt.
\newblock Realistic modeling of strongly correlated electron systems: An introduction to the lda+ dmft approach.
\newblock {\em International Journal of Modern Physics B}, 15(19n20):2611--2625, 2001.

\bibitem{sun2002extended}
Ping Sun and Gabriel Kotliar.
\newblock Extended dynamical mean-field theory and gw method.
\newblock {\em Physical Review B}, 66(8):085120, 2002.

\bibitem{boehnke2016strong}
Lewin Boehnke, Fredrik Nilsson, Ferdi Aryasetiawan, and Philipp Werner.
\newblock When strong correlations become weak: Consistent merging of gw and dmft.
\newblock {\em Physical Review B}, 94(20):201106, 2016.

\bibitem{zhu2021ab}
Tianyu Zhu and Garnet Kin-Lic Chan.
\newblock Ab initio full cell gw+ dmft for correlated materials.
\newblock {\em Physical Review X}, 11(2):021006, 2021.

\bibitem{gull2011continuous}
Emanuel Gull, Andrew~J Millis, Alexander~I Lichtenstein, Alexey~N Rubtsov, Matthias Troyer, and Philipp Werner.
\newblock Continuous-time monte carlo methods for quantum impurity models.
\newblock {\em Reviews of Modern Physics}, 83(2):349--404, 2011.

\bibitem{vidberg1977solving}
HJ~Vidberg and JW~Serene.
\newblock Solving the eliashberg equations by means of n-point pad{\'e} approximants.
\newblock {\em Journal of Low Temperature Physics}, 29:179--192, 1977.

\bibitem{jiani2021nevanlinna}
Jiani Fei, Chia-Nan Yeh, and Emanuel Gull.
\newblock Nevanlinna analytical continuation.
\newblock {\em Phys. Rev. Lett.}, 126:056402, Feb 2021.

\bibitem{rozenberg1994metal}
Marcelo~J Rozenberg, Goetz Moeller, and Gabriel Kotliar.
\newblock The metal--insulator transition in the hubbard model at zero temperature ii.
\newblock {\em Modern Physics Letters B}, 8(08n09):535--543, 1994.

\bibitem{caffarel1994exact}
Michel Caffarel and Werner Krauth.
\newblock Exact diagonalization approach to correlated fermions in infinite dimensions: Mott transition and superconductivity.
\newblock {\em Physical review letters}, 72(10):1545, 1994.

\bibitem{zgid2012truncated}
Dominika Zgid, Emanuel Gull, and Garnet Kin-Lic Chan.
\newblock Truncated configuration interaction expansions as solvers for correlated quantum impurity models and dynamical mean-field theory.
\newblock {\em Physical Review B—Condensed Matter and Materials Physics}, 86(16):165128, 2012.

\bibitem{lin2013efficient}
Chungwei Lin and Alexander~A Demkov.
\newblock Efficient variational approach to the impurity problem and its application to the dynamical mean-field theory.
\newblock {\em Physical Review B—Condensed Matter and Materials Physics}, 88(3):035123, 2013.

\bibitem{ganahl2015efficient}
Martin Ganahl, Markus Aichhorn, Hans~Gerd Evertz, Patrik Thunstr{\"o}m, Karsten Held, and Frank Verstraete.
\newblock Efficient dmft impurity solver using real-time dynamics with matrix product states.
\newblock {\em Physical Review B}, 92(15):155132, 2015.

\bibitem{bauernfeind2017fork}
Daniel Bauernfeind, Manuel Zingl, Robert Triebl, Markus Aichhorn, and Hans~Gerd Evertz.
\newblock Fork tensor-product states: Efficient multiorbital real-time dmft solver.
\newblock {\em Physical Review X}, 7(3):031013, 2017.

\bibitem{cao2021tree}
Xiaodong Cao, Yi~Lu, P~Hansmann, and Maurits~W Haverkort.
\newblock Tree tensor-network real-time multiorbital impurity solver: Spin-orbit coupling and correlation functions in sr 2 ruo 4.
\newblock {\em Physical Review B}, 104(11):115119, 2021.

\bibitem{cao2024finite}
Xiaodong Cao, E~Miles Stoudenmire, and Olivier Parcollet.
\newblock Finite-temperature minimally entangled typical thermal states impurity solver.
\newblock {\em Physical Review B}, 109(24):245113, 2024.

\bibitem{knizia2012density}
Gerald Knizia and Garnet Kin-Lic Chan.
\newblock Density matrix embedding: A simple alternative to dynamical mean-field theory.
\newblock {\em Physical review letters}, 109(18):186404, 2012.

\bibitem{knizia_2013}
Gerald Knizia and Garnet Kin-Lic Chan.
\newblock Density matrix embedding: A strong-coupling quantum embedding theory.
\newblock {\em Journal of chemical theory and computation}, 9(3):1428--1432, 2013.

\bibitem{wouters_2016}
Sebastian Wouters, Carlos~A Jim{\'e}nez-Hoyos, Qiming Sun, and Garnet K-L Chan.
\newblock A practical guide to density matrix embedding theory in quantum chemistry.
\newblock {\em Journal of chemical theory and computation}, 12(6):2706--2719, 2016.

\bibitem{lanata2012efficient}
Nicola Lanata, Hugo~UR Strand, Xi~Dai, and Bo~Hellsing.
\newblock Efficient implementation of the gutzwiller variational method.
\newblock {\em Physical Review B—Condensed Matter and Materials Physics}, 85(3):035133, 2012.

\bibitem{lanata2017critical}
Nicola Lanata, Tsung-Han Lee, Yong-Xin Yao, Vladan Stevanovi{\'c}, and Vladimir Dobrosavljevi{\'c}.
\newblock Critical role of electronic correlations in determining crystal structure of transition metal compounds.
\newblock {\em arXiv preprint arXiv:1710.08586}, 2017.

\bibitem{lanata2019connection}
Nicola Lanat{\`a}, Tsung-Han Lee, Yong-Xin Yao, Vladan Stevanovi{\'c}, and Vladimir Dobrosavljevi{\'c}.
\newblock Connection between mott physics and crystal structure in a series of transition metal binary compounds.
\newblock {\em npj Computational Materials}, 5(1):30, 2019.

\bibitem{ayral2017dynamical}
Thomas Ayral, Tsung-Han Lee, and Gabriel Kotliar.
\newblock Dynamical mean-field theory, density-matrix embedding theory, and rotationally invariant slave bosons: A unified perspective.
\newblock {\em Physical Review B}, 96(23):235139, 2017.

\bibitem{lee2023accuracy_single}
Tsung-Han Lee, Nicola Lanat{\`a}, and Gabriel Kotliar.
\newblock Accuracy of ghost rotationally invariant slave-boson and dynamical mean field theory as a function of the impurity-model bath size.
\newblock {\em Physical Review B}, 107(12):L121104, 2023.

\bibitem{lanata2023derivation}
Nicola Lanat{\`a}.
\newblock Derivation of the ghost gutzwiller approximation from quantum embedding principles: Ghost density matrix embedding theory.
\newblock {\em Physical Review B}, 108(23):235112, 2023.

\bibitem{sriluckshmy_2021}
P.~V. Sriluckshmy, Max Nusspickel, Edoardo Fertitta, and George~H. Booth.
\newblock Fully algebraic and self-consistent effective dynamics in a static quantum embedding.
\newblock {\em Phys. Rev. B}, 103:085131, Feb 2021.

\bibitem{lanata2022operatorial}
Nicola Lanat{\`a}.
\newblock Operatorial formulation of the ghost rotationally invariant slave-boson theory.
\newblock {\em Physical Review B}, 105(4):045111, 2022.

\bibitem{lee2023accuracy_multi}
Tsung-Han Lee, Corey Melnick, Ran Adler, Nicola Lanat{\`a}, and Gabriel Kotliar.
\newblock Accuracy of ghost-rotationally-invariant slave-boson theory for multiorbital hubbard models and realistic materials.
\newblock {\em Physical Review B}, 108(24):245147, 2023.

\bibitem{mejuto2023efficient}
Carlos Mejuto-Zaera and Michele Fabrizio.
\newblock Efficient computational screening of strongly correlated materials: Multiorbital phenomenology within the ghost gutzwiller approximation.
\newblock {\em Physical Review B}, 107(23):235150, 2023.

\bibitem{lee2024charge}
Tsung-Han Lee, Corey Melnick, Ran Adler, Xue Sun, Yongxin Yao, Nicola Lanat{\`a}, and Gabriel Kotliar.
\newblock Charge self-consistent density functional theory plus ghost rotationally invariant slave-boson theory for correlated materials.
\newblock {\em Physical Review B}, 110(11):115126, 2024.

\bibitem{frank2024active}
Marius~S Frank, Denis~G Artiukhin, Tsung-Han Lee, Yongxin Yao, Kipton Barros, Ove Christiansen, and Nicola Lanat{\`a}.
\newblock Active learning approach to simulations of strongly correlated matter with the ghost gutzwiller approximation.
\newblock {\em Physical Review Research}, 6(1):013242, 2024.

\bibitem{tagliente2025revealing}
Antonio~Maria Tagliente, Carlos Mejuto-Zaera, and Michele Fabrizio.
\newblock Revealing spinons by proximity effect.
\newblock {\em Physical Review B}, 111(12):125110, 2025.

\bibitem{mejuto2024quantum}
Carlos Mejuto-Zaera.
\newblock Quantum embedding for molecules using auxiliary particles--the ghost gutzwiller ansatz.
\newblock {\em Faraday Discussions}, 254:653--681, 2024.

\bibitem{white1992density}
Steven~R White.
\newblock Density matrix formulation for quantum renormalization groups.
\newblock {\em Physical review letters}, 69(19):2863, 1992.

\bibitem{zhai2023block2}
Huanchen Zhai, Henrik~R Larsson, Seunghoon Lee, Zhi-Hao Cui, Tianyu Zhu, Chong Sun, Linqing Peng, Ruojing Peng, Ke~Liao, Johannes T{\"o}lle, et~al.
\newblock Block2: A comprehensive open source framework to develop and apply state-of-the-art dmrg algorithms in electronic structure and beyond.
\newblock {\em The Journal of Chemical Physics}, 159(23), 2023.

\bibitem{huron1973iterative}
B~Huron, J~P Malrieu, and P~Rancurel.
\newblock Iterative perturbation calculations of ground and excited state energies from multiconfigurational zeroth-order wavefunctions.
\newblock {\em J. Chem. Phys., v. 58, no. 12, pp. 5745-5759}, 06 1973.

\bibitem{ohtsuka2017selected}
Yuhki Ohtsuka and Jun-ya Hasegawa.
\newblock Selected configuration interaction method using sampled first-order corrections to wave functions.
\newblock {\em The Journal of Chemical Physics}, 147(3), 2017.

\bibitem{levine2020casscf}
Daniel~S Levine, Diptarka Hait, Norm~M Tubman, Susi Lehtola, K~Birgitta Whaley, and Martin Head-Gordon.
\newblock Casscf with extremely large active spaces using the adaptive sampling configuration interaction method.
\newblock {\em Journal of chemical theory and computation}, 16(4):2340--2354, 2020.

\bibitem{kanno2023quantum}
Keita Kanno, Masaya Kohda, Ryosuke Imai, Sho Koh, Kosuke Mitarai, Wataru Mizukami, and Yuya~O Nakagawa.
\newblock Quantum-selected configuration interaction: Classical diagonalization of hamiltonians in subspaces selected by quantum computers.
\newblock {\em arXiv preprint arXiv:2302.11320}, 2023.

\bibitem{nakagawa2024adapt}
Yuya~O Nakagawa, Masahiko Kamoshita, Wataru Mizukami, Shotaro Sudo, and Yu-ya Ohnishi.
\newblock Adapt-qsci: Adaptive construction of an input state for quantum-selected configuration interaction.
\newblock {\em Journal of Chemical Theory and Computation}, 20(24):10817--10825, 2024.

\bibitem{mikkelsen2024quantum}
Mathias Mikkelsen and Yuya~O Nakagawa.
\newblock Quantum-selected configuration interaction with time-evolved state.
\newblock {\em arXiv preprint arXiv:2412.13839}, 2024.

\bibitem{ieva2025quantum}
Ieva Liepuoniute, Kirstin~D. Doney, Javier Robledo~Moreno, Joshua~A. Job, William~S. Friend, and Gavin~O. Jones.
\newblock Quantum-centric computational study of methylene singlet and triplet states.
\newblock {\em Journal of Chemical Theory and Computation}, 0(0):null, 0.
\newblock PMID: 40357738.

\bibitem{yu2025sample}
Jeffery Yu, Javier Robledo~Moreno, Joseph Iosue, Luke Bertels, Daniel Claudino, Bryce Fuller, Peter Groszkowski, Travis~S Humble, Petar Jurcevic, William Kirby, et~al.
\newblock Sample-based krylov quantum diagonalization.
\newblock {\em arXiv e-prints}, pages arXiv--2501, 2025.

\bibitem{motta2023bridging}
Mario Motta, Kevin~J Sung, K~Birgitta Whaley, Martin Head-Gordon, and James Shee.
\newblock Bridging physical intuition and hardware efficiency for correlated electronic states: the local unitary cluster jastrow ansatz for electronic structure.
\newblock {\em Chemical Science}, 14(40):11213--11227, 2023.

\bibitem{abdurakhimov2024technology}
Leonid Abdurakhimov, Janos Adam, Hasnain Ahmad, Olli Ahonen, Manuel Algaba, Guillermo Alonso, Ville Bergholm, Rohit Beriwal, Matthias Beuerle, Clinton Bockstiegel, et~al.
\newblock Technology and performance benchmarks of iqm's 20-qubit quantum computer.
\newblock {\em arXiv preprint arXiv:2408.12433}, 2024.

\bibitem{peng2020simulating}
Tianyi Peng, Aram~W Harrow, Maris Ozols, and Xiaodi Wu.
\newblock Simulating large quantum circuits on a small quantum computer.
\newblock {\em Physical review letters}, 125(15):150504, 2020.

\bibitem{lowe2023fast}
Angus Lowe, Matija Medvidovi{\'c}, Anthony Hayes, Lee~J O'Riordan, Thomas~R Bromley, Juan~Miguel Arrazola, and Nathan Killoran.
\newblock Fast quantum circuit cutting with randomized measurements.
\newblock {\em Quantum}, 7:934, 2023.

\bibitem{eckstein2005hopping}
Martin Eckstein, Marcus Kollar, Krzysztof Byczuk, and Dieter Vollhardt.
\newblock Hopping on the bethe lattice: Exact results for densities of states and dynamical mean-field theory.
\newblock {\em Phys. Rev. B}, 71:235119, Jun 2005.

\bibitem{frank2021quantum}
Marius~S. Frank, Tsung-Han Lee, Gargee Bhattacharyya, Pak Ki~Henry Tsang, Victor~L. Quito, Vladimir Dobrosavljevi\ifmmode~\acute{c}\else \'{c}\fi{}, Ove Christiansen, and Nicola Lanat\`a.
\newblock Quantum embedding description of the anderson lattice model with the ghost gutzwiller approximation.
\newblock {\em Phys. Rev. B}, 104:L081103, Aug 2021.

\bibitem{gao2024distributed}
Hong Gao, Satoshi Imamura, Akihiko Kasagi, and Eiji Yoshida.
\newblock Distributed implementation of full configuration interaction for one trillion determinants.
\newblock {\em Journal of Chemical Theory and Computation}, 20(3):1185--1192, 2024.

\bibitem{holmes2016heat}
Adam~A Holmes, Norm~M Tubman, and CJ~Umrigar.
\newblock Heat-bath configuration interaction: An efficient selected configuration interaction algorithm inspired by heat-bath sampling.
\newblock {\em Journal of chemical theory and computation}, 12(8):3674--3680, 2016.

\bibitem{evangelisti1983convergence}
Stefano Evangelisti, Jean-Pierre Daudey, and Jean-Paul Malrieu.
\newblock Convergence of an improved cipsi algorithm.
\newblock {\em Chemical Physics}, 75(1):91--102, 1983.

\bibitem{loos2020the}
Pierre-François Loos, Yann Damour, and Anthony Scemama.
\newblock The performance of cipsi on the ground state electronic energy of benzene.
\newblock {\em The Journal of Chemical Physics}, 153(17):176101, 11 2020.

\bibitem{reinholdt2025exposing}
Peter Reinholdt, Karl~Michael Ziems, Erik~Rosendahl Kjellgren, Sonia Coriani, Stephan Sauer, and Jacob Kongsted.
\newblock Exposing a fatal flaw in sample-based quantum diagonalization methods.
\newblock {\em arXiv preprint arXiv:2501.07231}, 2025.

\bibitem{motta2024quantum}
Mario Motta, Kevin~J Sung, and James Shee.
\newblock Quantum algorithms for the variational optimization of correlated electronic states with stochastic reconfiguration and the linear method.
\newblock {\em The Journal of Physical Chemistry A}, 128(40):8762--8776, 2024.

\bibitem{bartlett2007coupled}
Rodney~J Bartlett and Monika Musia{\l}.
\newblock Coupled-cluster theory in quantum chemistry.
\newblock {\em Reviews of Modern Physics}, 79(1):291--352, 2007.

\bibitem{mitarai2021constructing}
Kosuke Mitarai and Keisuke Fujii.
\newblock Constructing a virtual two-qubit gate by sampling single-qubit operations.
\newblock {\em New Journal of Physics}, 23(2):023021, 2021.

\bibitem{brenner2023optimal}
Lukas Brenner, Christophe Piveteau, and David Sutter.
\newblock Optimal wire cutting with classical communication.
\newblock {\em arXiv preprint arXiv:2302.03366}, 2023.

\bibitem{piveteau2022quasiprobability}
Christophe Piveteau, David Sutter, and Stefan Woerner.
\newblock Quasiprobability decompositions with reduced sampling overhead.
\newblock {\em npj Quantum Information}, 8(1):12, 2022.

\bibitem{pashayan2015estimating}
Hakop Pashayan, Joel~J Wallman, and Stephen~D Bartlett.
\newblock Estimating outcome probabilities of quantum circuits using quasiprobabilities.
\newblock {\em Physical review letters}, 115(7):070501, 2015.

\bibitem{harada2024doubly}
Hiroyuki Harada, Kaito Wada, and Naoki Yamamoto.
\newblock Doubly optimal parallel wire cutting without ancilla qubits.
\newblock {\em PRX Quantum}, 5:040308, Oct 2024.

\bibitem{ibm}
Circuit cutting.
\newblock https://docs.quantum.ibm.com/guides/qiskit-addons-cutting.

\bibitem{bechtold2023investigating}
Marvin Bechtold, Johanna Barzen, Frank Leymann, Alexander Mandl, Julian Obst, Felix Truger, and Benjamin Weder.
\newblock Investigating the effect of circuit cutting in qaoa for the maxcut problem on nisq devices.
\newblock {\em Quantum Science and Technology}, 8(4):045022, 2023.

\bibitem{ufrecht2024optimal}
Christian Ufrecht, Laura~S Herzog, Daniel~D Scherer, Maniraman Periyasamy, Sebastian Rietsch, Axel Plinge, and Christopher Mutschler.
\newblock Optimal joint cutting of two-qubit rotation gates.
\newblock {\em Physical Review A}, 109(5):052440, 2024.

\bibitem{lu2019natural}
Y~Lu, X~Cao, P~Hansmann, and MW~Haverkort.
\newblock Natural-orbital impurity solver and projection approach for green's functions.
\newblock {\em Physical Review B}, 100(11):115134, 2019.

\bibitem{kohn2021efficient}
Lucas Kohn and Giuseppe~E Santoro.
\newblock Efficient mapping for anderson impurity problems with matrix product states.
\newblock {\em Physical Review B}, 104(1):014303, 2021.

\bibitem{peruzzo2014variational}
Alberto Peruzzo, Jarrod McClean, Peter Shadbolt, Man-Hong Yung, Xiao-Qi Zhou, Peter~J Love, Al{\'a}n Aspuru-Guzik, and Jeremy~L O’brien.
\newblock A variational eigenvalue solver on a photonic quantum processor.
\newblock {\em Nature communications}, 5(1):4213, 2014.

\bibitem{Cai_2023_QEMreview}
Zhenyu Cai, Ryan Babbush, Simon~C. Benjamin, Suguru Endo, William~J. Huggins, Ying Li, Jarrod~R. McClean, and Thomas~E. O'Brien.
\newblock Quantum error mitigation.
\newblock {\em Rev. Mod. Phys.}, 95:045005, Dec 2023.

\bibitem{spall1998overview}
James~C Spall.
\newblock An overview of the simultaneous perturbation method for efficient optimization.
\newblock {\em Johns Hopkins apl technical digest}, 19(4):482--492, 1998.

\bibitem{guerci2023time}
Daniele Guerci, Massimo Capone, and Nicola Lanat{\`a}.
\newblock Time-dependent ghost gutzwiller nonequilibrium dynamics.
\newblock {\em Physical Review Research}, 5(3):L032023, 2023.

\bibitem{schmitt2025cutting}
Lukas Schmitt, Christophe Piveteau, and David Sutter.
\newblock Cutting circuits with multiple two-qubit unitaries.
\newblock {\em Quantum}, 9:1634, 2025.

\bibitem{gentinetta2024overhead}
Gian Gentinetta, Friederike Metz, and Giuseppe Carleo.
\newblock Overhead-constrained circuit knitting for variational quantum dynamics.
\newblock {\em Quantum}, 8:1296, 2024.

\bibitem{liu2025quantum}
Kecheng Liu and Zhenyu Cai.
\newblock Quantum error mitigation for sampling algorithms.
\newblock {\em arXiv preprint arXiv:2502.11285}, 2025.

\bibitem{anuar2024operator}
Aeishah~Ameera Anuar, Fran{\c{c}}ois Jamet, Fabio Gironella, Fedor Simkovic~IV, and Riccardo Rossi.
\newblock Operator-projected variational quantum imaginary time evolution.
\newblock {\em arXiv preprint arXiv:2409.12018}, 2024.

\bibitem{gluza2024double}
Marek Gluza, Jeongrak Son, Bi~Hong Tiang, Yudai Suzuki, Zo{\"e} Holmes, and Nelly~HY Ng.
\newblock Double-bracket quantum algorithms for quantum imaginary-time evolution.
\newblock {\em arXiv preprint arXiv:2412.04554}, 2024.

\bibitem{larocca2025barren}
Martin Larocca, Supanut Thanasilp, Samson Wang, Kunal Sharma, Jacob Biamonte, Patrick~J Coles, Lukasz Cincio, Jarrod~R McClean, Zo{\"e} Holmes, and M~Cerezo.
\newblock Barren plateaus in variational quantum computing.
\newblock {\em Nature Reviews Physics}, pages 1--16, 2025.

\bibitem{miller2025simulation}
Aaron Miller, Zo{\"e} Holmes, {\"O}zlem Salehi, Rahul Chakraborty, Anton Nyk{\"a}nen, Zolt{\'a}n Zimbor{\'a}s, Adam Glos, and Guillermo Garc{\'\i}a-P{\'e}rez.
\newblock Simulation of fermionic circuits using majorana propagation.
\newblock {\em arXiv preprint arXiv:2503.18939}, 2025.

\bibitem{weinberg2017quspin}
Phillip Weinberg and Marin Bukov.
\newblock Quspin: a python package for dynamics and exact diagonalisation of quantum many body systems part i: spin chains.
\newblock {\em SciPost Physics}, 2(1):003, 2017.

\bibitem{sun2020recent}
Qiming Sun, Xing Zhang, Samragni Banerjee, Peng Bao, Marc Barbry, Nick~S Blunt, Nikolay~A Bogdanov, George~H Booth, Jia Chen, Zhi-Hao Cui, et~al.
\newblock Recent developments in the pyscf program package.
\newblock {\em The Journal of Chemical Physics}, 153(2), 2020.

\bibitem{ffsim}
{The ffsim developers}.
\newblock {ffsim: Faster simulations of fermionic quantum circuits.}

\bibitem{chen2025quantum}
I~Chen, Aleksei Khindanov, Carlos Salazar, Humberto~Munoz Barona, Feng Zhang, Cai-Zhuang Wang, Thomas Iadecola, Nicola Lanat{\`a}, and Yong-Xin Yao.
\newblock Quantum-classical embedding via ghost gutzwiller approximation for enhanced simulations of correlated electron systems.
\newblock {\em arXiv preprint arXiv:2506.01204}, 2025.

\end{thebibliography}

\appendix

\section{Errors in higher moments of the DOS}
\label{appendix:moments}
\begin{figure}
\centering
\hspace{0.2cm}
\includegraphics[width=0.41\textwidth]{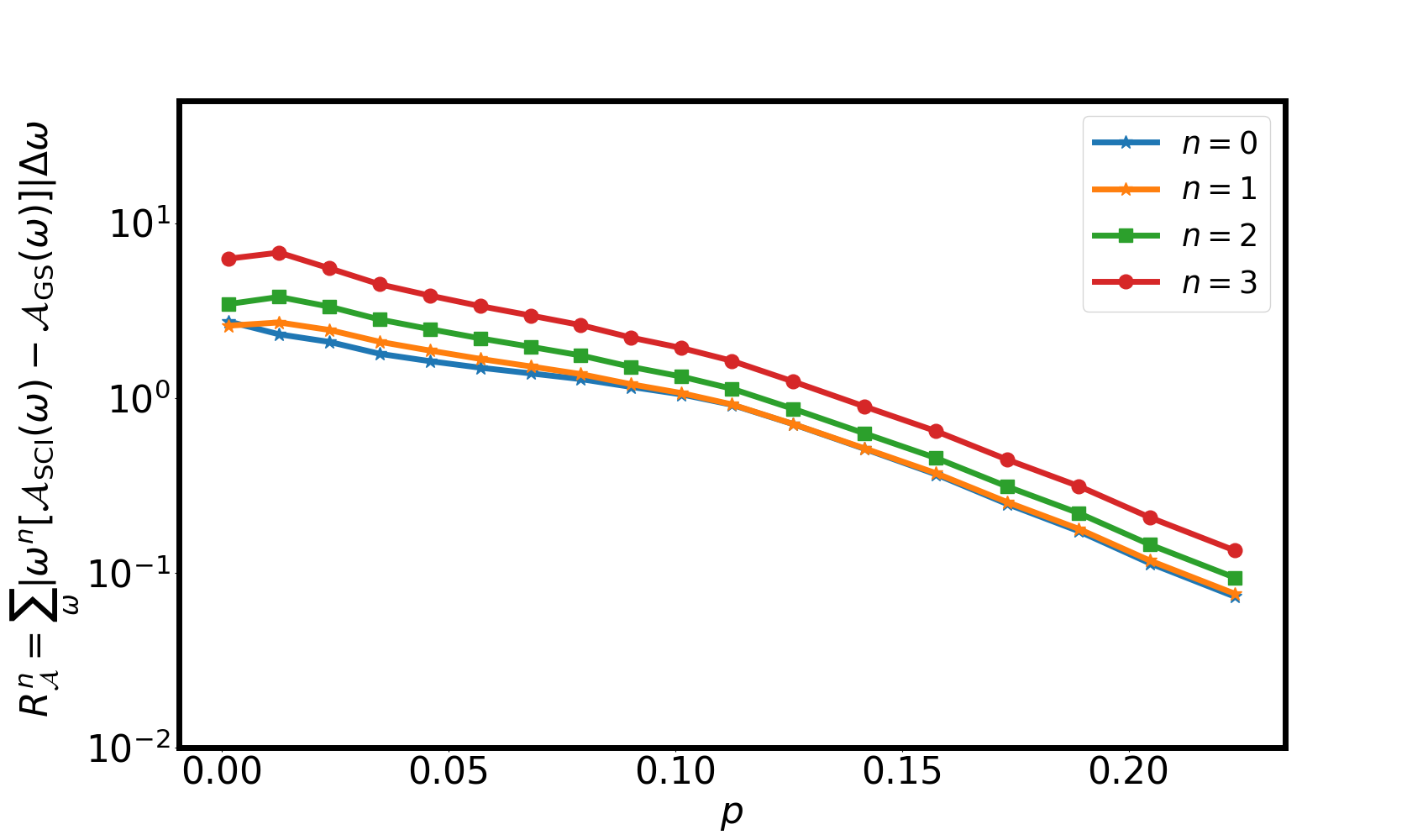}
\caption{\justifying Comparison of the error $R_{\mathcal{A}}^n$ (defined by Eq.\eqref{eq:moments}) for different moments $n$ of the density of states for $U=2$, $N_g=9$ and as a function of the fraction of the total CI states in the SCI basis, $p$.}
\label{fig:gfmom}
\end{figure}

A systematic comparison of the structural features of the DOS can be obtained by comparing the errors incurred on its higher moments, $R_{\mathcal{A}}^n$ , which can be calculated using:
\begin{align}
    R_{\mathcal{A}}^n & = \sum_\omega |\omega^n [\mathcal{A}_{\text{SCI}} (\omega) - \mathcal{A}_{\text{GS}} (\omega) ]| \Delta\omega. 
    \label{eq:moments}
\end{align}
In Fig.~\ref{fig:gfmom}, the error for the first few moments of the DOS are compared as a function of the truncation in size of the CI basis set given by the fraction parameter $p = |\mathcal{S}_{\text{SCI}}|/|\mathcal{S}_{\text{sym}}|$ (see main text). The error for all moments decreases at a the same pace, revealing that same, up to a small constant prefactor, computational resources are required to recover the higher-moment features of the DOS to a similar degree of accuracy.

\begin{figure}
\centering
\hspace{-1cm}
\includegraphics[width=0.4\textwidth]{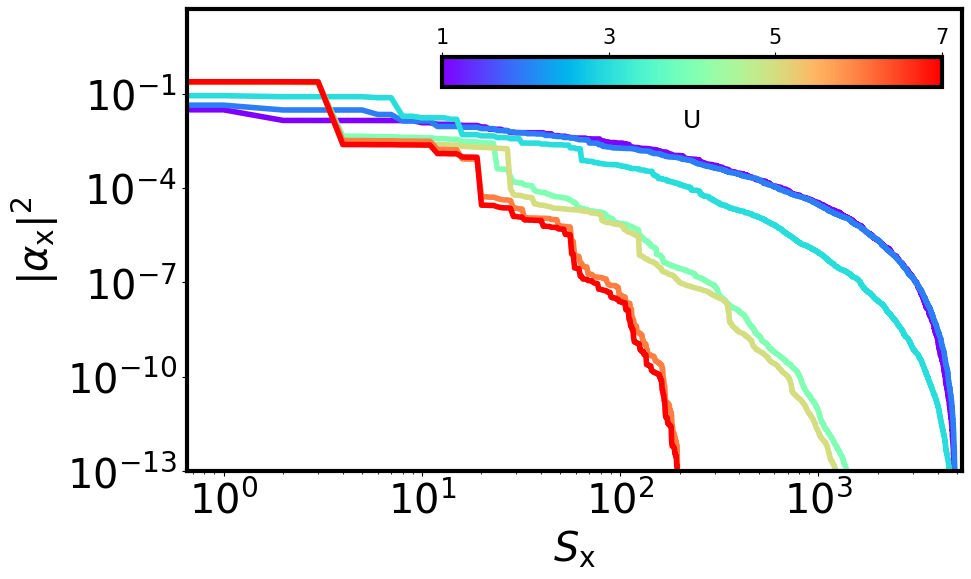}
\caption{\justifying The rate of decay of CI weights as a function of U for $5$ rotated orbitals. Similar trends are obtained for other rotated Hamiltonian bases.}
\label{fig:diffrotudependence}
\end{figure}
\begin{figure*}[ht!]
\centering
\includegraphics[width=0.407\textwidth]{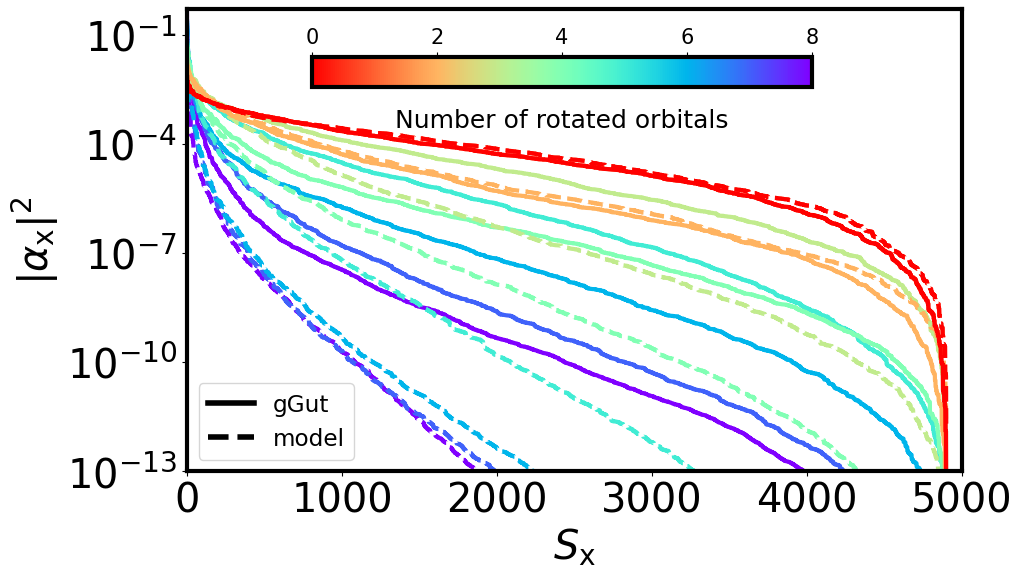}
\includegraphics[width=0.4\textwidth]{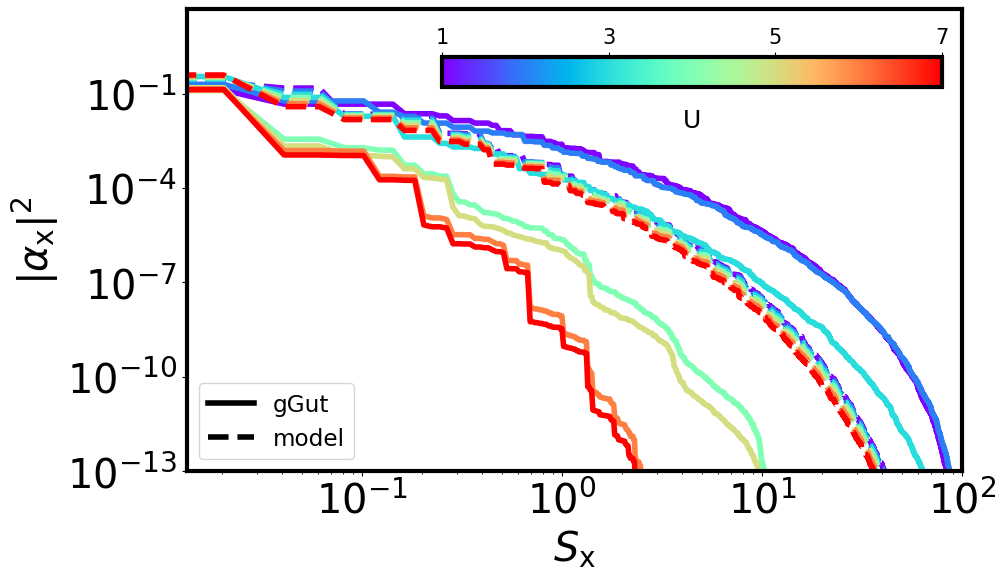}
\caption{\justifying 
(a) A comparison of the weights $|\alpha|^2$ of CIs $S_x$, ordered by their magnitude from the FCI ground state wavefunction at $U=2$, $N_g = 7$ is shown for different rotated Hamiltonian bases (see main text). Results for converged gGut embedding Hamiltonians (solid lines) are compared to a generic single-band Anderson impurity model with parameters set to match those of Ref.\cite{yu2025sample} (dashed lines). (b) For the rotation indexed $7$ and called star configuration, a comparison of the weights of CIs in the FCI ground state wavefunction of the aforementioned two types of embedding Hamiltonians are shown for different interaction strengths $1\leq U \leq 7$ at $N_g=7$.}
\label{fig:aim_compare_model}
\end{figure*}

\begin{figure*}[ht!]
\centering
\begin{subfigure}{0.59\textwidth}
\includegraphics[width=\textwidth]{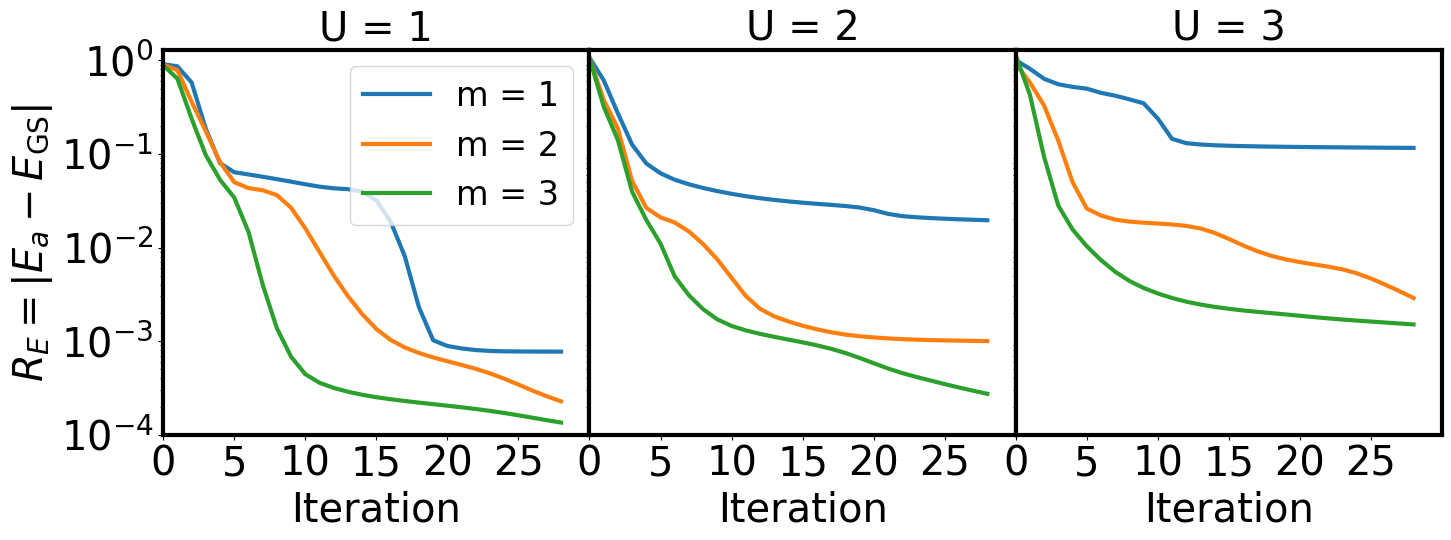}
    \caption{} \label{fig:8a}
  \end{subfigure}%
  \begin{subfigure}{0.347\textwidth}
  \includegraphics[width=\textwidth]{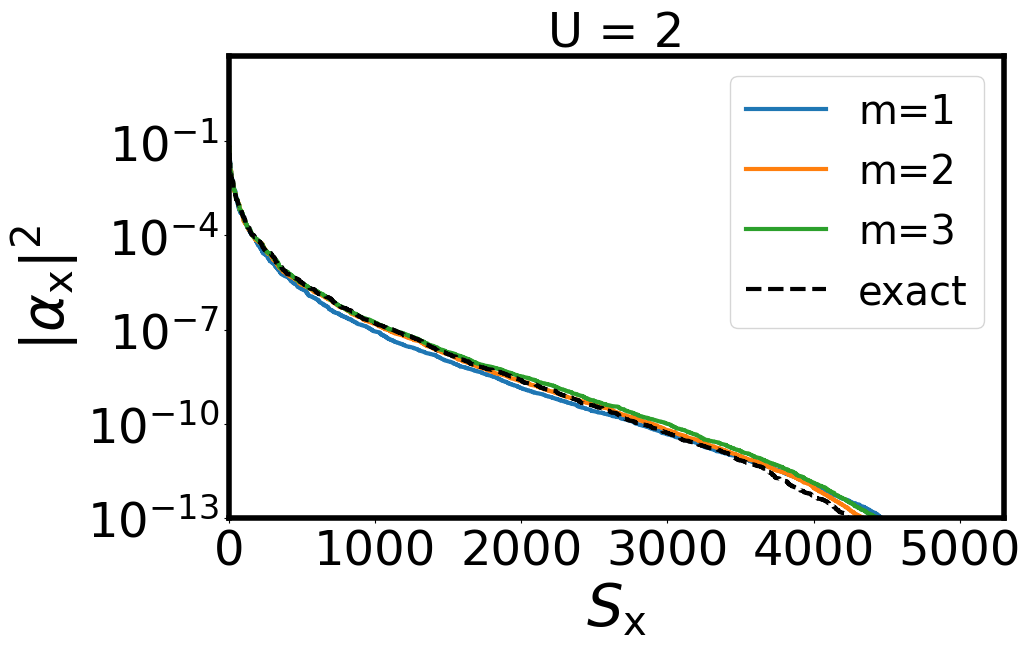}
    \caption{} \label{fig:8b}
  \end{subfigure}%
\caption{\justifying (a) Convergence of the optimized  LUCJ ansatz energy $E_a$ to the ground state energy $E_{\text{GS}}$ as a function of the optimized iteration. Results are shown for different numbers of ansatz layers $m=\{1,2,3\}$ with $N_g=7$ and $U=\{1,2,3\}$. (b) Comparison of the CI weights, ordered by their magnitude, from the optimized LUCJ ansatz wavefunctions (solid lines) with different numbers of ansatz layers $m=\{1,2,3\}$ and compared to the exact ground state (dashed line).}
\label{fig:optlucj}
\end{figure*}

\begin{figure*}
\centering
\includegraphics[width=0.9\textwidth]{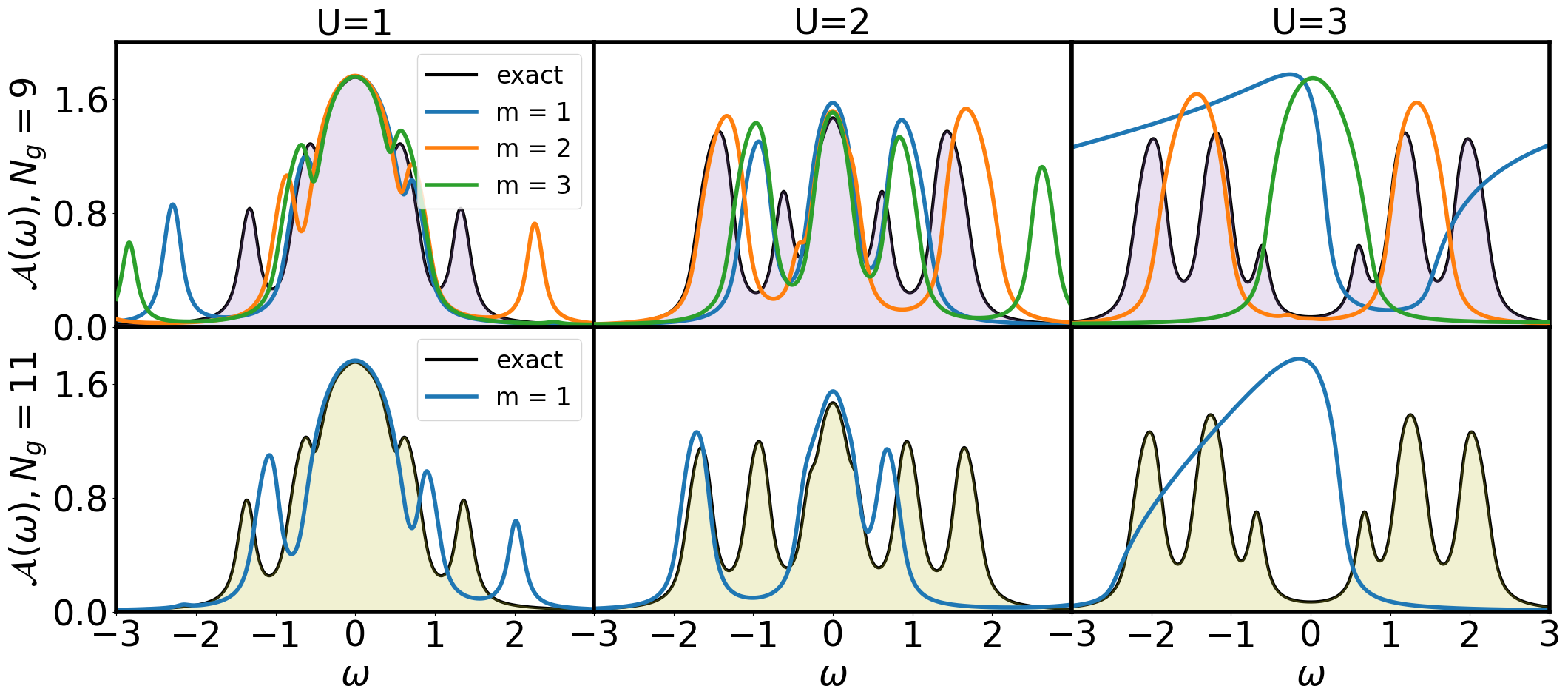}
\caption{\justifying Comparison of the approximate ground state DOS obtained from the optimized simulated LUCJ ansatz circuits with different numbers of ansatz layers, $m=\{1,2,3\}$, for interaction strengths $U=\{1,2,3\}$ and $N_g=\{9,11\}$. Results are benchmarked against the exact DOS obtained from FCI (filled out in color). The number of parameters per ansatz layer grow as $m N(N+2)/4$ where $N$ is the system size.}
\label{figapp:gf_lucj_ansatz_add}
\end{figure*}

\begin{figure*}
\centering
\includegraphics[width=0.9\textwidth]{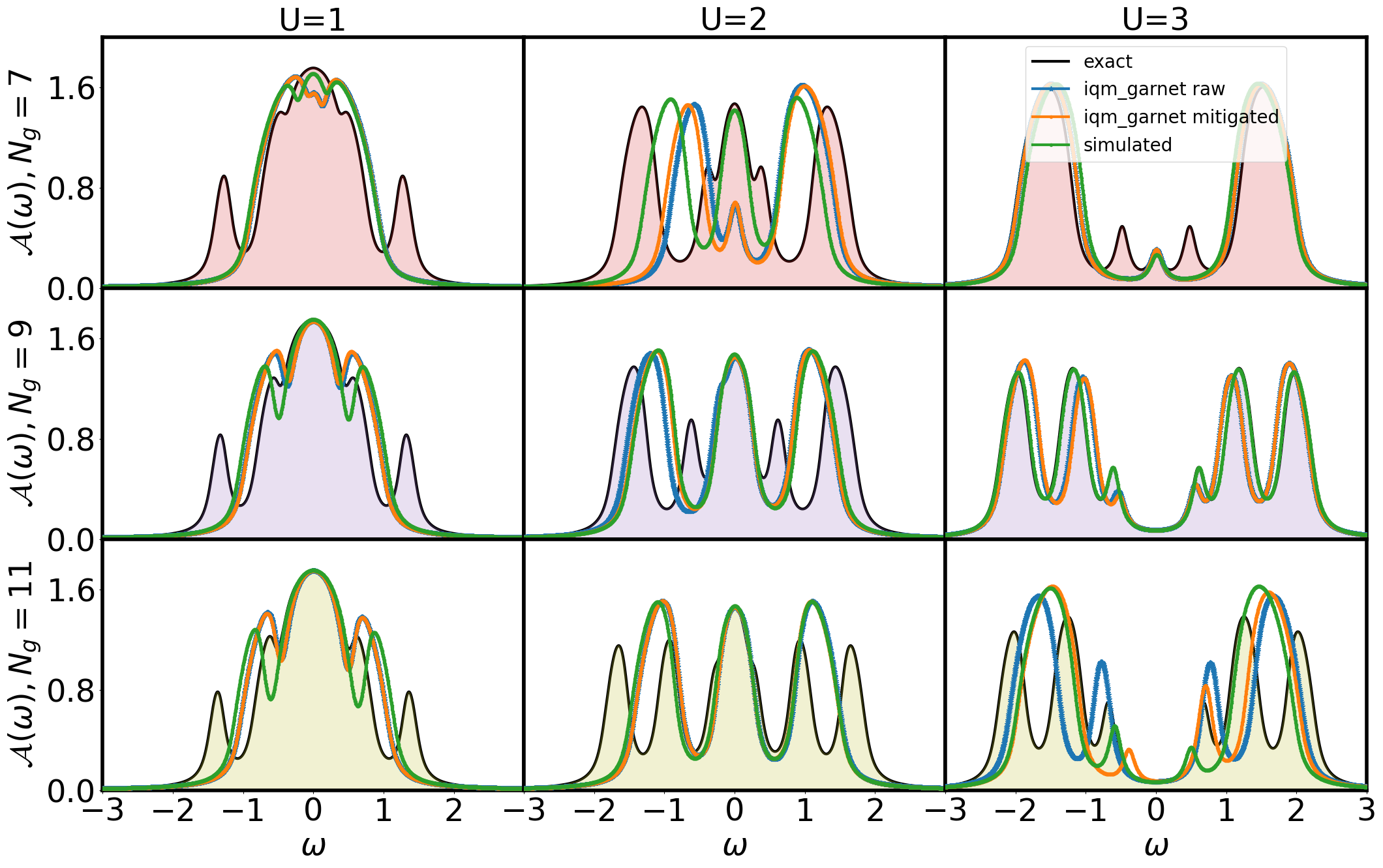}
\caption{\justifying The DOS obtained from QSCI constructed with a total of $3.2 \cdot 10^5$ samples ($2 \cdot 10^4$ per sub-circuit) obtained from an LUCJ ansatz circuit with $m=1$ layers and two wire cuts, and prepared on IQM Garnet quantum hardware. Raw results after post-selection based on symmetry verification are compared to results that have been post-processed using readout error mitigation\cite{Cai_2023_QEMreview}. Results are shows as a function of interaction strength $U=\{1,2,3\}$, the bath sizes $N_g = \{7,9,11\}$ and the respective QSCI basis truncation fractions $p=\{0.05,0.05,0.01\}$. Hardware results are benchmarked against noiseless simulations using the same ansatz circuits and number of shots, as well as against the exact DOS obtained from FCI (filled out in color).}
\label{fig:rawmitig}
\end{figure*}

\section{Details of the classical gGut calculations}
\label{appendix:details}
For the classical gGut calculations, we use a momentum  of $\mathbf{k} = (k_x,k_y) = (20,20)$   (Eq.~\eqref{eq:qphamiltonian}) and set the temperature to $T=0.002$ for the Fermi-Dirac distribution function used in Eq.~\eqref{eq:rhoqp} and Eq.~\eqref{eq:Deltaqp}. A small broadening of $\delta = 0.01$ is applied to compute the Green's function in Eq.~\eqref{eq:greensfunc}, which is then modified to the form:
\begin{align}
    G (\mathbf{k},\omega) = \sum_{ab} \Omega_{a}^\dagger \left[ \omega + i\delta - \Omega\epsilon_{\mathbf{k}} \Omega^\dagger -\Lambda^{\text{qp}} \right]_{ab}^{-1} \Omega_{b}.
\end{align}

The number of CI basis states in the correct symmetry sector $\mathcal{S}_{\text{sym}}$ is tabulated below for different bath sizes $N_g$:\\

\begin{center}
\begin{tabular}{||c | c||} 
 \hline
 $N_g$ & $|\mathcal{S}_{\text{sym}}|$ \\ [0.5ex] 
 \hline\hline
 1 & 4 \\ 
 \hline
 3 & 36  \\
 \hline
 5 & 400  \\
 \hline
 7 & 4900  \\
 \hline
 9 & 63504 \\
 \hline
 11 & 853776 \\ [1ex] 
 \hline
\end{tabular}
\end{center}

For a small bath size, the classical gGut calculation converges within $10$ loop iterations. We found that gGut is particularly difficult to converge for $N_g=11$ ghosts at $U=3$ and used the trick of warm-starting from previously converged calculations at other values of $U$ to help the convergence. For LUCJ, with parameters starting from $T_2$ amplitudes obtained from a CCSD calculation \cite{bartlett2007coupled} for a $U$ close to the phase transition did not always provide good convergence and instead small angles for the rotational gates were used as a starting point for the SPSA optimization. In total, $4$ macro-iterations with $10$ micro iterations of the SPSA were executed for every loop iteration of gGut.

\section{CI decay rate dependence on the interaction strength} \label{appendix:udependence}
In the main text, we have investigated the ground state of the embedding Hamiltonian rotated into the star basis as a function of interaction strength $U$ and found that the sparseness of such Hamiltonians increases as the interaction strength grows. In fact, we found this trend to be true also for other rotated Hamiltonian bases and in Fig.~\ref{fig:diffrotudependence} we provide another example for the basis with $5$ rotated bath orbitals, where the aforementioned trend can be clearly observed.

\section{Rotation of the Basis}
\label{appendix:rotation}

In the main text, we performed an analysis of the optimal single-particle Hamiltonian rotation for embedding Hamiltonians obtained from gGut. Here, we compare those results to applying rotations to a generic single-band Anderson impurity model, whose Hamiltonian has the form:
\begin{align}
    \hat{\mathcal{H}}^{\text{AIM}} &=  U  \hat{n}_{0\uparrow}^{d} \hat{n}_{0\downarrow}^{d} + \mu \sum_{\sigma} \hat{n}_{0\sigma}^{d}  
    +\Delta_1 \sum_{\sigma}  \left(\hat{d}^{\dagger}_{0\sigma} \hat{d}^{\phantom{\dagger}}_{1\sigma}
    + \hat{d}^{\dagger}_{1\sigma} \nonumber \hat{d}^{\phantom{\dagger}}_{0\sigma} \right) \\&
    -  \sum_{ \sigma \in \{\uparrow,\downarrow\}} \sum_{a=1}^{N_g -1}  \Lambda_{a,a+1} \left[\hat{d}^{\dagger}_{a,\sigma}\hat{d}^{\phantom{\dagger}}_{a+1,\sigma} + \hat{d}^{\dagger}_{a+1,\sigma}\hat{d}^{\phantom{\dagger}}_{a,\sigma}\right],  
\end{align}
with the impurity creation and annihilation operators $\hat{d}^{\dagger}_{0\sigma}$/ $\hat{d}_{0\sigma}$ interacting with the corresponding ladder operators defined on the bath sites $\hat{d}^{\dagger}_{a\sigma}$/$\hat{d}_{a\sigma}$, where $a \in \{1,\cdots N_g\}$. A comparison of this generic model to those obtained from a converged self-consistent gGut loop calculation is presented in  Fig.~\ref{fig:aim_compare_model}. The model parameters are chosen to be $\Lambda_{a,a+1} = - \Delta_1 = 1$ and $\mu = -U/2$, matching those presented in Ref.~\cite{yu2025sample}. As is observed from Fig.~\ref{fig:aim_compare_model}a for $U=2$ and $N_g=7$, the best rotation for the generic model, similar to the case of gGut, is the canonical basis. The star configuration indexed as $7$, however remains the most practical choice considering additional the computational costs (see main text). Comparing the two types of embedding Hamiltonian it also becomes clear that for the given interaction strength of $U=2$, the ground state of the generic Hamiltonian is considerably more sparse than the one obtained from gGut, regardless of the choice of Hamiltonian rotation with the notable exception of the original unrotated basis indexed $0$ and the chain configuration indexed $1$. 

In Fig.~\ref{fig:aim_compare_model}b, we investigate the two embedding Hamiltonians rotated to the star configuration basis (indexed $7$) as a function of the interaction strength $U$. Naturally, the decay rate of CI weights does not depend on the values of $U$ for the generic model, which shows that such a model is not able to capture the features of a self-consistent embedding calculation like gGut. We also see that, depending on the value of $U$, this can lead to a more or less sparse ground state of the generic embedding Hamiltonian as compared to the one obtained from gGut.
\section{Further discussions on the LUCJ ansatz}
\label{appendix:lucjmore}

The training of the LUCJ ansatz reduced to the symmetry sector shows that the error in the energy $R_E=|E_{a}-E_{\text{GS}}|$ drops by roughly a factor of $10$ per additional ansatz layer for $N_g=7$, as shown Fig.~\ref{fig:optlucj}a. For lower $U$, a single layer is sufficient to obtain an accuracy of $10^{-3}$. The same accuracy is, however, not achieved at higher $U$ even with $m=3$ layers, showing a high dependence of the optimized ansatz properties on the interaction strength in the embedding Hamiltonian. Fig.~\ref{fig:optlucj}b shows that the decay rate of the CI weights, obtained from optimized LUCJ wavefunctions essentially does not depend on the number of ansatz layer, $m$. The DOS obtained by preparing the pretrained LUCJ ansatz for $N_g=7$ ghosts are shown in the main text while the ones corresponding to $N_g=9$ and $N_g=11$ ghosts are presented in Fig.~\ref{figapp:gf_lucj_ansatz_add}. We observe that the quality of the LUCJ ansatz further worsens as the bath size and interaction strengths are increased.

Since the number of parameters grows as $m N(N+2)/4$ where $N$ is the system size, it becomes increasingly difficult to obtain convergence in the energy within a fixed number of iterations and in a reasonable computational time. Therefore, for $N_g=11$ the calculations are restricted to $m=1$, which is reflected by the poor accuracy of the computed DOS.

\section{REM results}
\label{appendix:rem}
Fig.~\ref{fig:rawmitig} shows the QSCI plus circuit cutting results for the density of states obtained from the IQM hardware. CIs are post-processed based on symmetry verification (blue, labeled raw) and additionally treated with a read-out error mitigation algorithm (orange, labeled mitigated) \cite{Cai_2023_QEMreview}. In most cases the results obtained for both versions of the post-processing are similar, showing that QSCI naturally exhibits a high degree of robustness to noise in the quantum hardware. We also include the results from the simulator with the same shot budget of $3.2 \cdot 10^5$ samples as our hardware runs (green, labeled sim). 

Even though a similar cutoff in the number of CIs included in the $\mathcal{S}_{\text{SCI}}$ set, is applied to the simulated and quantum hardware data, much fewer CIs are actually obtained for the simulated case. In particular, for $U=3$ only a fraction $p=0.02$ of the total CI set is obtained for $N_g=9$ and an even smaller fraction is recovered for $p=0.005$ for $N_g=11$. Despite that, for $N_g=9$, $U=3$, the complete DOS is already reproduced using the simulated results.
There is also a marked difference when using circuit cutting as compared to the results from the uncut circuit, as shown in Fig.~\ref{fig:nocut}. The simulated results enforces the need for better error mitigation techniques dedicated to sampling the computational states from the hardware.

\end{document}